\documentclass[showpacs,prb,aps,floatfix,twocolumn,superscriptaddress]{revtex4-1}

\usepackage[pdftex]{graphicx}
\usepackage{dcolumn}% Align table columns on decimal point
\usepackage{bm}% bold math
\usepackage{amsmath}
\usepackage{array}
\usepackage{color}
\usepackage{float}
\usepackage{subfigure}
\usepackage{txfonts}
\usepackage{wasysym}
\usepackage{multirow}
\usepackage{nicefrac}
\usepackage{sidecap}
\usepackage{xcolor,cancel}
\usepackage[normalem]{ulem}
\usepackage{mathtools}

\newcommand{\dlangle}{\langle\langle}

\newcommand{\drangle}{\rangle\rangle}

\newcommand{\e}{\varepsilon}

\newcommand{\up}{\uparrow}
\newcommand{\down}{\downarrow}

\newcommand{\veck}{\mathbf{k}}

 \newcommand{\greenfunc}[2]{{\dlangle{#1};}{{#2}\drangle}}

\newcommand{\tip}{ {\rm tip}}
\newcommand{\imp}{ {\rm imp}}
\newcommand{\M}{ {\rm M}}
\usepackage{hyperref}
\hypersetup{
    colorlinks,%
    citecolor=blue,%
    linkcolor=blue,%
    urlcolor=blue
}

\begin{document}

\title{Reentrant Kondo effect in a quantum impurity coupled to a metal-semiconductor hybrid contact}

\author{G.~Diniz}
\affiliation{Instituto de F\'isica, Universidade Federal de Uberl\^andia, 
Uberl\^andia, Minas Gerais 38400-902, Brazil.}

\author{G.~S.~Diniz}

\affiliation{Curso de F\'isica, Universidade Federal de Jata\'i, Jata\'i, GO 75801-615, Brazil.}

\author{G.~B.~Martins}
\affiliation{Instituto de F\'isica, Universidade Federal de Uberl\^andia, 
Uberl\^andia, Minas Gerais 38400-902, Brazil.}
\email[Corresponding author: ]{gbmartins@ufu.br}

\author{E.~Vernek}
\affiliation{Instituto de F\'isica, Universidade Federal de 
Uberl\^andia, Uberl\^andia, Minas Gerais 38400-902, Brazil.}

\affiliation{Department of Physics and Astronomy, and Nanoscale and Quantum
Phenomena Institute, Ohio University, Athens, Ohio 45701-2979, USA.}

\date{\today}

\begin{abstract}
	Using the Numerical Renormalization Group (NRG) and Anderson's poor man's scaling, we show that a system containing 
	a quantum impurity (QI), strongly coupled to a semiconductor (with gap $2 \Delta$) and weakly coupled to a metal, displays 
	a \emph{reentrant} Kondo stage as one gradually lowers the temperature T. The NRG analysis of the corresponding 
	Single Impurity Anderson Model (SIAM), through the impurity's thermodynamic and spectral properties, 
	shows that the reentrant stage is characterized by a second sequence of SIAM fixed points, viz., free orbital 
	(FO) $\rightarrow$ local moment (LM) $\rightarrow$ strong coupling (SC). In the higher temperature stage, the SC 
	fixed point (with a Kondo temperature $T_{K1}$) is unstable, while the lower temperature Kondo screening exhibits 
	a much lower  Kondo temperature $T_{K2}$, associated to a stable SC fixed point. The results clearly indicate that 
	the reentrant Kondo screening is associated to an effective SIAM, with an effective Hubbard repulsion $U_{\rm eff}$, 
	whose value is clearly identifiable in the impurity's local density of states. This low temperature effective SIAM, 
	which we dub as \emph{reentrant} SIAM, behaves as a \emph{replica} of the high temperature 
	(bare) SIAM. 
	The second stage RG flow (obtained through NRG), whose FO fixed point emerges for $T \approx \Delta < T_{K1}$, 
	takes over once the RG flows away from the unstable first stage SC fixed point. The intuitive picture that 
	emerges from our analysis is that the first Kondo state develops through impurity screening by semiconducting electrons, 
	while the second Kondo state involves screening by metallic electrons, once the semiconducting electrons are out of 
	reach to thermal excitations ($T < \Delta$) and only the metallic (low) spectral weight 
	inside the gap is available for impurity screening. This switch implies that the first Kondo cloud is much smaller than 
	the second, since the NRG results show that, for all parameter ranges analyzed, $T_{K2} \ll T_{K1}$. 
	Last, but not least, we analyze a hybrid system formed by a QI `sandwiched' between an armchair graphene 
	nanoribbon (AGNR) and a scanning tunneling microscope (STM) tip (an AGNR+QI+STM system), with respective couplings set 
	to reproduce the generic model described above. 
	The energy gap ($2 \Delta$) in the AGNR can be externally tuned by an electric-field-induced Rashba spin-orbit interaction. 
	We analyzed this system for realistic parameter values, using NRG, and concluded that the reentrant SIAM, with its 
	associated second 
	stage Kondo, is worthy of experimental investigation. 

\end{abstract}
\maketitle

\section{Introduction}\label{sec:intro}

Understanding the low-temperature physics of a many-body interacting system is always a challenging task. Despite the simple form of the mutual interaction between pairs of its constituents, such a system, collectively, ofttimes behaves in an unexpected manner. Indeed, this beautiful aspect of nature has been insightfully discussed in a seminal paper by P.~W.~Anderson~\cite{Anderson393}. Within this context, the archetypal example, in condensed matter physics, is that of the ground state of the many-body Kondo problem~\cite{Kondo1964,Hewson-Kondo}.  

The Kondo physics of a single magnetic impurity coupled to a metallic host is a well-understood problem~\cite{Hewson-Kondo}, which can be experimentally studied in detail by coupling a quantum dot (QD) to a metallic contact~\cite{Goldhaber-Gordon1998}, while its essential physical 
properties are captured by the well-known single impurity Anderson model (SIAM)~\cite{Anderson1961}. A renormalization-group (RG) 
analysis of the SIAM~\cite{Krishna-murthy1980} shows that the system crosses over three different fixed points as the temperature is lowered: (i) the unstable free orbital (FO) fixed point, in which the impurity is effectively decoupled from the conduction band, (ii) the also unstable local moment (LM) fixed point, where the impurity acquires a highly fluctuating magnetic moment, and (iii) the stable strong coupling (SC) fixed point, in which the magnetic moment of the impurity becomes fully screened by the conduction band electrons. The characteristic temperature below which the impurity moment is screened is the so-called Kondo temperature, $T_K$. The SIAM, so to speak, provides a rich, although the simplest, description of the Kondo physics in QDs. The scenario 
presented above provides a generic picture of the physics of the SIAM, which remains qualitatively valid whenever the density of states of the conduction electrons exhibits no special features close to the Fermi level. Richer Kondo physics can be found if the conduction band exhibits structures such as a pseudo-gap or zero-energy peaks, like van-Hove singularities. These features have been studied in great detail by several authors~\cite{Bulla2008}.

An interesting, but less studied situation, is the case in which the conduction band is that of a semiconductor, i.e, a spectra characterized by a finite gap $\Delta$. The richness of the Kondo physics resulting from the interplay between $T_K$ and $\Delta$ has been studied since almost three decades ago using a variety of numerical and analytical techniques, for instance: Quantum Monte Carlo (QMC), by Takegahara \emph{et al.}~\cite{Takegahara1992,Takegahara1993} and T.~Saso~\cite{Saso1992}, 
poor man's scaling (PMS), $1/N$ expansion, non-crossing approximation (NCA) and QMC, by Ogura and Saso~\cite{Ogura1993}, 
using Green's function, within equation-of-motion techniques, plus Hartree-Fock, by Cruz \emph{et al.}~\cite{Cruz1995}, density 
matrix renormalization group (DMRG), by Yu and Guerrero~\cite{Yu1996}, numerical renormalization group (NRG), by Takegahara \emph{et al.}~\cite{Takegahara1992,Takegahara1993} and  Chen and  Jayaprakash~\cite{Chen1998}, Density Matrix NRG (DM-NRG), by Moca and Roman~\cite{Moca2010}, as well as perturbation theory and the local-moment approach, by Galpin and Logan~\cite{Galpin2008a,Galpin2008b}. 
\begin{figure}[t!]\centering
\subfigure{\includegraphics[clip,width=3.3in]{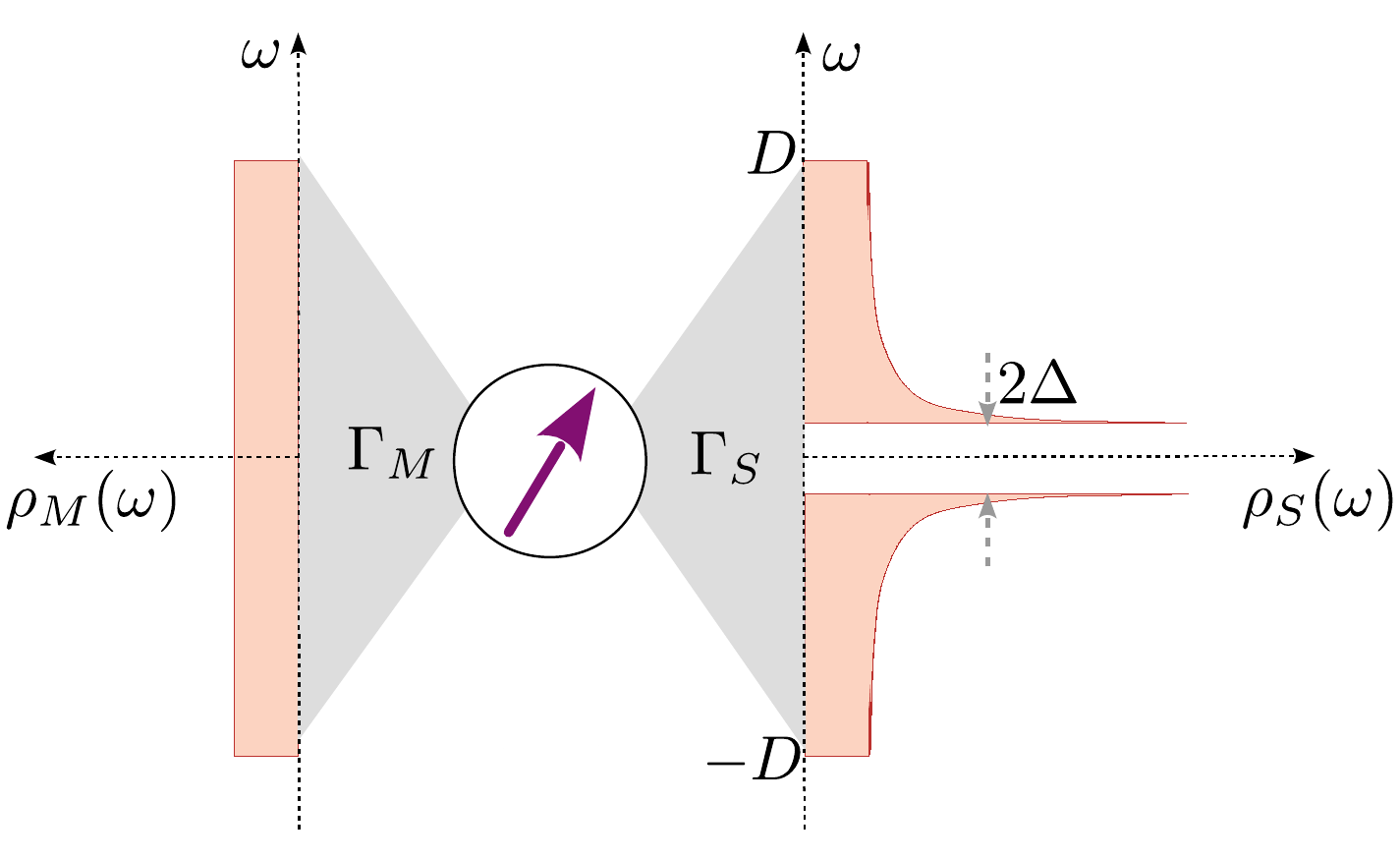}}
\caption{Schematic representation of a QD coupled to a
	metallic lead (left) and to a semiconducting lead (right). The metallic lead is represented by a 
	flat density of states $\rho_M(\omega)$, while the semiconducting lead is modeled by an 
	energy dependent density of states $\rho_S(\omega)$ characterized by a 
	gap $2 \Delta$. $D$ is a cutoff energy and represents the bandwidth 
	of conduction electrons and is taken as our energy unit.} 
\label{fig1_new}
\end{figure} 

The earliest results pointed to the existence of a Kondo ground state (a SC fixed point) whenever $\Delta < \Delta_c$, where the critical gap $\Delta_c$ should fulfill the relation $\Delta_c \lesssim T_K$, being $T_K$ defined as the Kondo temperature for $\Delta=0$. However, NRG results~\cite{Takegahara1992,Chen1998,Moca2010} 
have indicated that a finite critical gap $\Delta_c$ only exists \emph{away} from half-filling, while at half-filling any arbitrarily small gap (i.e., any $\Delta > 0$) results in the ground state becoming a 
doublet, i.e., switching from the standard Kondo-singlet SC fixed point (for $\Delta=0$) to a doublet LM fixed point. This qualitative difference (half-filling vs.~away-from-half-filling) has been confirmed by analytical calculations~\cite{Galpin2008a} and the local-moment approach~\cite{Galpin2008b}, where it was shown that the ground state away from half-filling is a so-called generalized Fermi liquid, while it is a non-Fermi liquid for all finite values of $\Delta$ at half-filling. In addition, DM-NRG calculations~\cite{Moca2010} studied the quantum phase transition (QPT) occurring away from half-filling for $\Delta = \Delta_c$ and showed the formation of a single bound state when the system 
is in the SC regime ($\Delta < \Delta_c$), and the formation of an additional 
one once the system transitions to the LM regime ($\Delta > \Delta_c$). 

In this work, we study two systems: the first is a slightly different model from the one already 
analyzed in the works described above, as it is composed of a QD [or a quantum impurity (QI)] 
that is \emph{strongly} coupled on the right to a semiconducting lead (with a gap $2 \Delta$) and on 
the left it is \emph{weakly} coupled to a \emph{metallic} lead (see Fig.~\ref{fig1_new}). The second system, which we 
believe to be a feasible experimental realization of the model just described, is based on a 
QI strongly coupled to an armchair graphene nanoribbon (AGNR), which is in an 
externally induced insulating phase \cite{bilayer}, and weakly coupled, through a small coupling $\Gamma_{\rm tip}$, 
to a scanning tunneling microscope (STM) tip (modeled as a \emph{metallic}-like band). 
This AGNR+QI+STM system is particularly attractive, as Kondo physics in 
carbon-based materials, mainly in \emph{bulk} samples, has attracted 
a great deal of attention in the last few years~\cite{Nygard,Jarillo,PhysRevB.77.045417,
PhysRevB.83.165449,0034-4885-76-3-032501,PhysRevB.84.165105,PhysRevB.88.201103,PhysRevB.87.075116,PhysRevB.90.035426,PhysRevB.100.115115}. 
The Kondo physics in graphene results from localized magnetic moments formed at 
vacancy sites~\cite{PhysRevB.83.241408,PhysRevB.88.075104,PhysRevB.97.155419,Jiang2018} 
or through the surface deposition of magnetic atoms \cite{Li_2013,nl501425n}, in which the 
local density of states may be modified by either disorder~\cite{Fuhrer,PhysRevB.90.201101} 
or by ripples induced 
by the underlying substrate~\cite{nl501425n}. Contrasting to the plethora of studies 
addressing the Kondo state in carbon nanotubes and on \emph{bulk} graphene, 
less attention has been devoted to this effect in nanoribbon 
systems~\cite{Busser2013,PhysRevB.89.035424,Li2017,PhysRevB.97.115444}. 
Depending on the shape of the edges of a graphene nanoribbon, 
either zigzag or armchair, its density of states near the Fermi level 
will be that of a semi-metal, for zigzag nanoribbons, owing to the remarkable existence of 
metallic states localized at its edges, or it could alternate between being 
semiconducting or metallic, for armchair nanoribbons, depending on its width~\cite{Wakabayashi2009}. 
Interesting Kondo physics can be exploited from graphene nanoribbons, 
as recently shown by Li et al.~\cite{Li2017}, which reported an unexpected Kondo resonance behavior in 
a magnetic-molecule/Au(111) coupled system, in which an AGNR was used as 
a \textit{bridge} to connect the molecule to the Au(111) surface, forming a hybrid 
structure. Their results showed that, thanks to their peculiar electronic 
properties, AGNRs were able to provide an effective coupling between the localized spin 
and the itinerant electrons in the Au(111) surface.

The main result in this work is that the PMS and NRG analysis, of the 
appropriate SIAM for modeling the first system mentioned in the preceding paragraph, reveals, 
as one lowers the temperature, a sequence of \emph{two} Kondo \emph{stages}. Both are characterized 
by the traditional sequence of SIAM fixed points (FO-LM-SC), where the higher temperature  
SC fixed point is unstable, with Kondo temperature $T_{K1}$, while the second stage has a 
stable SC fixed point with a much lower Kondo temperature $T_{K2}$. We dub 
the lower-temperature Kondo-state as a `\emph{reentrant} Kondo state', which is associated 
to an `emergent' effective SIAM, with an effective Hubbard $U_{\rm eff}$, in contrast to the 
`bare' SIAM associated to the first stage Kondo effect. The AGNR+QI+STM system, on the other hand, 
is a `real life' system where we claim, supported by NRG results for realistic 
parameters, the reentrant Kondo state may be experimentally observable.

The general organization of this work is as follows. In Sec.~\ref{sec:model} we 
present the SIAM that describes the first system and the specific parameter values used. 
For the sake of completeness, in Sec.~\ref{Sec:PMS} we disconnect the QI from 
the metallic band (keeping its coupling just to the semiconductor) and present 
a preliminary analysis, using Anderson's PMS~\cite{Anderson1970,Hewson-Kondo}, highlighting the 
interesting interplay between $T_K$ and $\Delta$. In Sec.~\ref{sec:reent}, we additionally (weakly) 
couple the magnetic impurity to the metallic band and 
study, through an analysis of the impurity thermodynamic properties, as well 
as its local density of states (LDOS), an interesting effect, the \emph{reentrant} Kondo effect, 
that can be briefly described as consisting of a sequence of two Kondo effects, where the Kondo temperature 
of the first ($T_{K1}$) is orders of magnitude higher that the second one ($T_{K2}$). Despite similarities 
with the so-called two-stage Kondo effect~\cite{Cornaglia2005}, there are important differences, 
the main one being that, in our system, the first Kondo effect is associated to an unstable 
SC fixed point, thus there is only one true Kondo state, which occurs below $T_{K2}$. 
In Sec.~\ref{sec:AGNR} we apply the ideas developed for the reentrant Kondo effect to a 
real system, viz., a QI coupled to an AGNR and an STM tip (see Fig.~\ref{fig1}). 
In Sec.~\ref{sec:conc} we present a summary of the results, together with our conclusions.

\section{Model and numerical results}
\label{sec:model}
The first system that we have studied is schematically described in Fig.~\ref{fig1_new}. 
In it, the semiconducting and the metallic density of states (DOS) seen by the QD 
are depicted to its right and left, respectively.  
As shown bellow, the presence of this metallic DOS will qualitatively change the many-body ground 
state of this system, in comparison to the ones analyzed in the literature, as described in the Introduction.

Thus, our model consists of an interacting QD coupled to a 
metallic lead, as well as to a semiconducting one (see Fig.~\ref{fig1_new}). 
This system is modelled by a Hamiltonian $H_{\rm SIAM}=H_{\rm imp}+H_{\rm S} +H_{\rm M}+H_{\rm Hyb}$, 
whose first term is given by
\begin{eqnarray}\label{H_imp}
H_{\rm imp} =\sum_{\sigma} \e_d d^\dagger_\sigma d_\sigma + U n_{d\up}n_{d\down}, 
\end{eqnarray}
where $d^\dagger_{\sigma}$ ($d_{\sigma}$) creates (annihilates) an electron 
with energy $\e_d$ and spin $\sigma=\uparrow \downarrow$ in the QD, 
$n_{d\sigma}=d^\dagger_{\sigma}d_\sigma$ is the QD occupancy, and $U$ 
represents the Coulomb interaction. The leads are described by 
\begin{eqnarray}\label{Anderson_Band}
H_{\rm S/M} =\sum_{\substack{\veck \sigma \\ a= {\rm S,M}} } \e_{a\veck}c^\dagger_{a\veck\sigma}c_{a\veck\sigma},
\end{eqnarray}
where $c^\dagger_{a\veck\sigma}$ ($c_{a\veck\sigma}$) creates (annihilates) an 
electron with momentum $\veck$, energy  $\e_{a\veck}$ and spin $\sigma$ in 
the metallic ($a={\rm M}$) or in the semiconducting ($a={\rm S}$) lead. 
Finally, the QD-leads hybridization is given by 
\begin{eqnarray}\label{Anderson_Hyb}
H_{\rm Hyb} =\sum_{\substack{ \veck \sigma \\ a={\rm S,M}} }\left(V_ {a\veck} d^\dagger_\sigma c_{a\veck\sigma}+{\rm H.c.} \right),
\end{eqnarray}
where $V_{a\veck}$ represents the hybridization matrix element that 
couples the impurity either to the metallic ($a={\rm M}$) or to 
the semiconducting ($a={\rm S}$) lead.
Here, we assume that the metallic lead is characterized by a 
flat DOS $\rho_M(\omega)=(1/2D)\Theta(D-|\omega|)$, where $D$ 
is the half band width ($\Theta$ is the Heaviside step function), 
while the semiconducting-lead DOS (schematically shown in 
Fig.~\ref{fig1_new}) is given by 
\begin{eqnarray}\label{DOS}
\rho_{S}(\omega) = \rho_0\frac{|\omega|}{\sqrt{{\omega}^2 - {\Delta}^2 }} \Theta(|\omega| - \Delta)\Theta(D-|\omega|).
\end{eqnarray}
Here, $2\Delta$ is the semiconducting gap and $\rho_0 = \frac{1}{2\sqrt{D^2-\Delta^2}}$ is a normalization factor. 
Assuming $V_{a\veck}\equiv V_{a}$ to be $\veck$-independent, for simplicity, the hybridization 
functions are defined as $\Gamma_a=\pi V_a^2\rho_a$ (for $a=S,M$). 

The Kondo physics in our model, for $\Gamma_{\rm S}=0$, corresponds to the traditional 
SIAM, which has been extensively studied over the last decades. In contrast, the 
situation where the QD couples solely to the semiconducting lead has received less 
attention (see the Introduction). Experimentally, the Kondo physics for magnetic impurities 
adsorbed in metallic surfaces has been studied through low-bias transport spectroscopy 
using an STM tip weakly coupled to the impurity. In our setup, the metallic lead 
serves not only to represent the STM tip, but also plays an important role in the 
NRG calculations, as it introduces a small, but finite, hybridization function 
at energies inside  the semiconducting gap $2\Delta$ (see Fig.~\ref{fig1_new}). 

In this work, we focus on the regime in which the QD is so weakly coupled to the metallic lead, 
in comparison to its coupling to the semiconducting lead ($\Gamma_{\rm M} \ll \Gamma_{\rm S}$), 
that any possible Kondo screening generated by conduction electrons in the metallic lead will occur at 
temperatures much lower than those associated to a possible Kondo screening occurring through electrons 
in the semiconducting lead. For our analysis in what follows, it is useful to 
define $\Gamma_0=\Gamma_{\rm M} + \Gamma_{\rm S} \approx \Gamma_{\rm S}$.

Note that all the calculations presented in this work, aside from those in Sec.~\ref{sec:AGNR}, 
where different parameters (when considered) are explicitly stated, 
were done for the following parameter values: $D=1$, the half-bandwidth, is our unit of energy, 
$U=0.5$ is the Coulomb repulsion for impurity double occupancy, the impurity 
energy level is set at the particle-hole-symmetric point $\e_d=-U/2$, and 
$\Gamma_0=0.05$. The NRG approach was performed using Wilson's discretization parameter 
set to $\Lambda$ = 2.5, 2000 many-body states were retained after each NRG iteration, 
and we made use of the $z$-trick averaging in the discretization procedure~\cite{Ljubljana}.

\subsection{Interplay between $T_K$ and $\Delta$: Effective Kondo Hamiltonian and scaling analysis \label{Sec:PMS}}

To reveal the intricate interplay between $T_K$~\cite{TK1} and $\Delta$, we will do a 
scaling analysis of the effective Kondo model, which can be derived from 
the SIAM by performing a Schrieffer-Wolff transformation~\cite{Schrieffer1966,Hewson-Kondo}. 
For now, we are solely interested in the impurity plus semiconductor subsystem, 
thus we set $V_{\rm M}=0$. The resulting Kondo model can be written as 
\begin{eqnarray}\label{Kondo}
H_{\rm K} &=& \sum_{\veck\sigma}\e_{S\veck\sigma} c^\dagger_{S\veck\sigma} 
c_{S\veck\sigma} + \sum_{\veck \veck^\prime}J_{S\veck\veck^\prime} \left[ 
S^z\left(c^\dagger_{S\veck \up}c_{S\veck^\prime\up}-c^\dagger_{S\veck 
\down}c_{S\veck^\prime\down} \right) \nonumber \right.\\
&& \left.+S^+c^\dagger_{S\veck\down}c_{S\veck^\prime\up}+
S^-c^\dagger_{S\veck\up}c_{S\veck^\prime\down}\right],
\end{eqnarray}
where $J_{S\veck\veck^\prime}$ is a Kondo coupling that can be written 
in terms of the SIAM parameters. For simplicity, we assume $V_{S\veck}$ 
to be $\veck$-independent and real, thus denoting it by $V_S$, resulting in  
$J_{S\veck\veck^\prime} \approx J_S =V_S^2 \left(\frac{1}{U+\e_d} - \frac{1}{\e_d} \right)$. 
(Note that, in what follows, for reasons that will be apparent soon, we will refer 
to $J_S$ as the bare coupling and denote it as $J_S^{(0)}$). In the above, we 
have neglected a scalar scattering potential, which in fact vanishes at the $\e_d=-U/2$ particle-hole 
symmetric point. Following Anderson's original idea~\cite{Anderson1970,Hewson-Kondo}, 
the scaling analysis consists of integrating out the degrees of freedom in the 
conduction band whose energies lie within the interval $[D-\delta D,D]$, for electrons, 
and $[-D,-D+\delta D]$, for holes, where $\delta D >0$. By doing so, we obtain 
an effective Kondo Hamiltonian where now the electrons are within a narrowed 
$\tilde D=D-\delta D$ conduction bandwidth, and with a renormalized coupling 
$\tilde J_S$, which obeys the scaling equation 
\begin{eqnarray}\label{E.Beta}
\frac{d\tilde J_S}{d (\ln\tilde{D})} &=& -2\rho_S(\tilde{D})\tilde J^2_S .
\end{eqnarray}
This equation has to be integrated from $D$ to some arbitrary energy 
$\tilde D< D$. Using Eq.~\eqref{DOS} for $\rho_S$, we obtain the general solution 
\begin{eqnarray}\label{J}
\frac{1}{\tilde{J}_S(\tilde D)} - \frac{1}{\tilde{J}_S(D)} &=& 2\rho_0 \left[\ln\left(\frac{\Delta}{ D+\sqrt{D^2-\Delta^2}}\right)
\Theta(\Delta-\tilde{D}) \nonumber \right.\\
&& \left.+ \ln\left(\frac{\tilde D +\sqrt{{\tilde D}^2-\Delta^2}}{ D+\sqrt{D^2-\Delta^2}} \right)\Theta(\tilde{D}-\Delta) \right], 
\end{eqnarray}
where $\tilde J_S(D)=J_S^{(0)}$ is the initial condition, which corresponds to 
(as mentioned above) the so-called bare Kondo coupling (i.e., the coupling before 
the rescaling of the conduction band).
As $\tilde D$ decreases, the expected SC fixed point is reached 
when  $J_S(\tilde D) \rightarrow \infty $. At this fixed point, the impurity and the 
conduction electrons form a many-body Kondo singlet. Within the PMS, the value of 
$D^*$, defined as $\tilde J_S(\tilde D=D^*) = \infty$, is identified with 
the Kondo temperature of the system.

The two terms inside the square brackets on the rhs of  Eq.~\eqref{J}, 
each multiplied to a different Heaviside step function, will thus be 
finite for different intervals of $\tilde D$: the first term for 
$\tilde D < \Delta$ and the second one for $\tilde D > \Delta$. 
This  implies, as we shall see, a qualitative change in the solutions 
when $\tilde D$ crosses $\Delta$. Starting with $\tilde D < \Delta$ 
(thus the second term vanishes), we obtain that 
\begin{eqnarray}
\frac{1}{\tilde{J_S}(\tilde D)} - \frac{1}{J_S^{(0)}} &=& 2\rho_0 \ln\left(\frac{\Delta}{ D+\sqrt{D^2-\Delta^2}}\right),  
\end{eqnarray}
which results in a finite, but constant, coupling $\tilde J_S(\tilde D)$, 
for any finite $\Delta$. Hence, no strong coupling fixed point [i.e., no divergence of $\tilde J_S(\tilde D)$] is expected.

On the other hand, the solution to Eq.~\eqref{J} for $\tilde D>\Delta$ 
(first term in Eq.~\eqref{J} vanishes), given by 
\begin{eqnarray}\label{J1}
\frac{1}{\tilde{J}_S(\tilde D)} - \frac{1}{J_{S}^{(0)}} &=& 2\rho_0 \ln\left(\frac{\tilde D +\sqrt{{\tilde D}^2-\Delta^2}}{ D+\sqrt{D^2-\Delta^2}}
\right),
\end{eqnarray}
allows for an infinite $\tilde J_S(\tilde D)$. Indeed, by setting $1/\tilde J_S(D^*) =0$ 
in Eq.~\eqref{J1}, after some algebraic manipulations we obtain that $D^*$ can be written as 
\begin{eqnarray}\label{TK}
D^* = \frac{1}{2}\left[\left({D+\sqrt{D^2-\Delta^2}}\right)e^{-g} + \frac{\Delta^2 }{D+\sqrt{D^2-\Delta^2}} e^{g} \right],
\end{eqnarray}
where $g= \left(2 \rho_0 J_S^{(0)}\right)^{-1}$. Obviously, $D^*$ is 
meaningful only if it lies within the interval $\Delta < D^* < D$. 
Upon imposing this condition on Eq.~\eqref{TK}, we find that, for a given $\Delta$, 
the bare coupling $J_S^{(0)}$  has to be larger than 
a critical $J_c$, given by~\cite{note0}
\begin{eqnarray}\label{JC}
\rho_0 J_c = \frac{1}{2} \left[\ln \left( \frac{D+\sqrt{D^2-\Delta^2}}{\Delta}\right) \right]^{-1}.
\end{eqnarray}
As mentioned in the Introduction, we know that this is an artifact of the poor 
man's scaling approach, since, at half filling, as shown through NRG and confirmed 
by other methods, there is no SC fixed point for any finite gap $\Delta$ in the 
semiconductor spectra. In the following, we will compare the critical coupling 
given by Eq.~\eqref{JC} with the numerical results obtained from 
NRG calculations for the corresponding Anderson model. To do so, 
it is convenient to express $J_c$ in terms of the Anderson model parameters. 
Defining $\Gamma_S^{(0)}=\pi V^2_S\rho_0$, we can write $J_S^{(0)}=4V_S^2/U=4\Gamma_S^{(0)}/(\pi \rho_0 U)$, 
at the particle-hole-symmetric point~\cite{rho0}. Thus, Eq.~\eqref{JC} can be rewritten as 
\begin{eqnarray}\label{Gamma_C}
\Gamma_c = \frac{\pi U}{8} \left[\ln \left( \frac{D+\sqrt{D^2-\Delta^2}}{\Delta}\right) \right]^{-1}.
\end{eqnarray}
\begin{figure}[ht!]
\centering
\subfigure{\includegraphics[clip,width=3.4in]{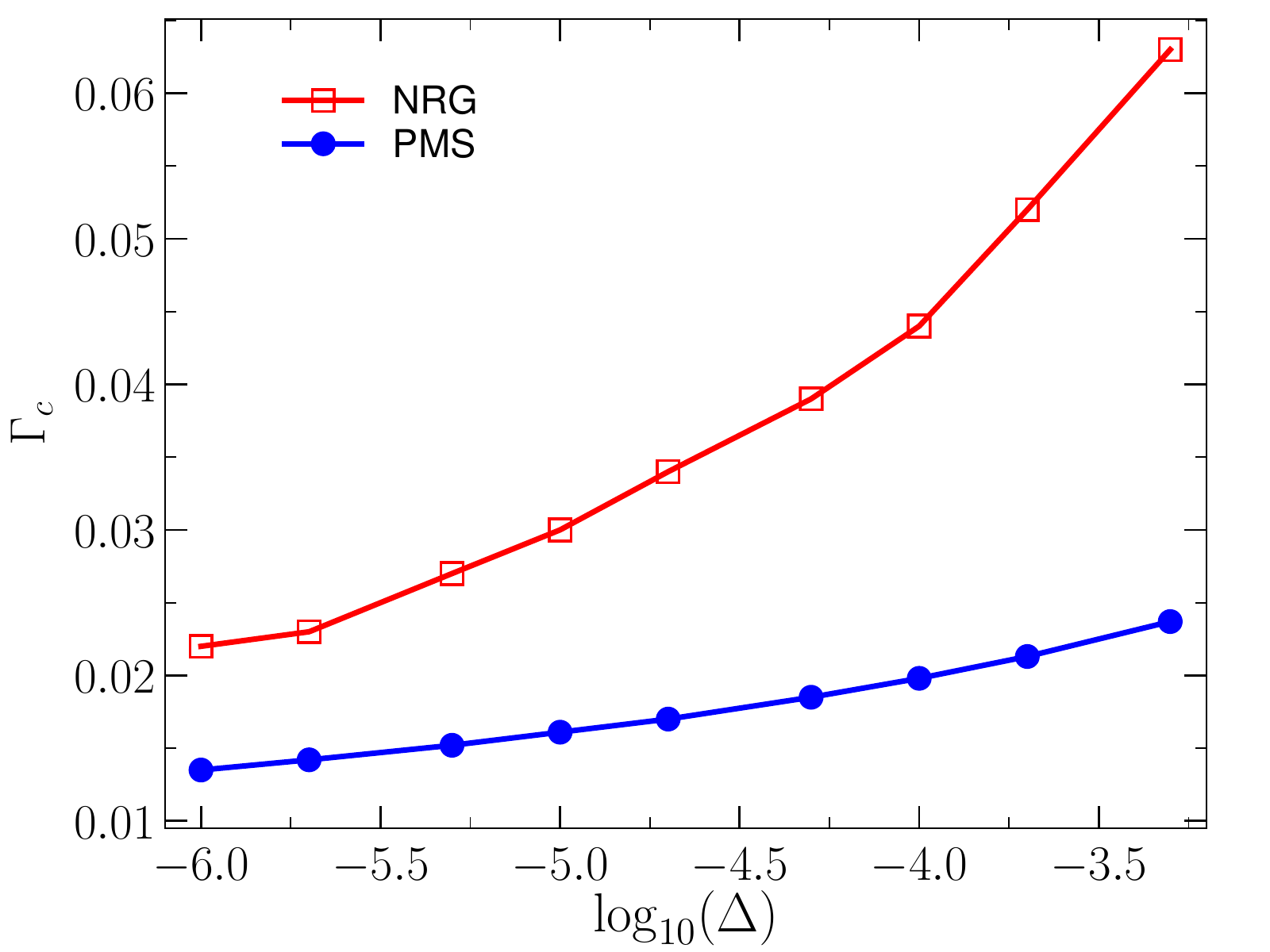}}
	\caption{$\Gamma_c$ obtained by PMS [(blue) dots], Eq.~\eqref{Gamma_C}, 
	and by NRG [(red) squares], as a function of $\Delta$ (in $\log$ scale). 
	The bare parameter values were $U=0.5$ and $\e_d=-0.25$.} 
\label{PMS}
\end{figure} 
In Fig.~\ref{PMS}, we plot $\Gamma_c$ vs $\Delta$ (in $\log$ scale) for $U=0.5$ and $\e_d=-0.25$, 
as obtained through the expression in Eq.~\eqref{Gamma_C} (blue dots) and compare it with the 
critical $\Gamma_{c}$ obtained by NRG (red squares). To determine whether 
there is a tendency to Kondo screening or not in the NRG calculations, we monitor the 
impurity magnetic moment $\mu^2_{\rm imp}(T)=k_BT 
\chi_{imp}(T)$ for decreasing temperature (not shown). Following Wilson's criterion~\cite{Hewson-Kondo}, 
we say that the Kondo screening takes place only if $\mu^2_{\rm imp}(T)$ becomes smaller 
than 0.07 as the system is cooled down. Thus, $\Gamma_c$ is defined as the smallest 
value of $\Gamma$, as obtained through NRG (red squares in Fig.~\ref{PMS}), 
for which this condition is still satisfied. It is interesting to notice that 
the $\Gamma_c$ obtained by NRG is systematically larger than the one obtained by PMS 
[Eq.~\eqref{Gamma_C}]. 
We note that there is a qualitative agreement between the PMS and NRG results, 
showing that $\Gamma_c$ increases with $\Delta$. This means that, as intuitively 
expected, a larger $\Delta$ requires stronger hybridization between the impurity 
and the (semiconducting) conduction electrons for the Kondo screening to take place. Last, but not 
least, taking into account that, as shown above, there is no SC fixed point for 
$\tilde{D} < \Delta$, the NRG results in Fig.~\ref{PMS} (red squares) do not describe 
the ground state of the $V_M=0$ Hamiltonian, but rather what we may call a 
finite-temperature-Kondo-phase (see below) associated to an unstable SC fixed 
point. As described in the Introduction, the ground state of the $V_M=0$ 
Hamiltonian corresponds to a doublet LM fixed point~\cite{Galpin2008a,Galpin2008b}. 

\subsection{Reentrant effective Anderson Hamiltonian}\label{sec:reent}
\begin{figure}[ht!]
\centering
\subfigure{\includegraphics[clip,width=3.4in]{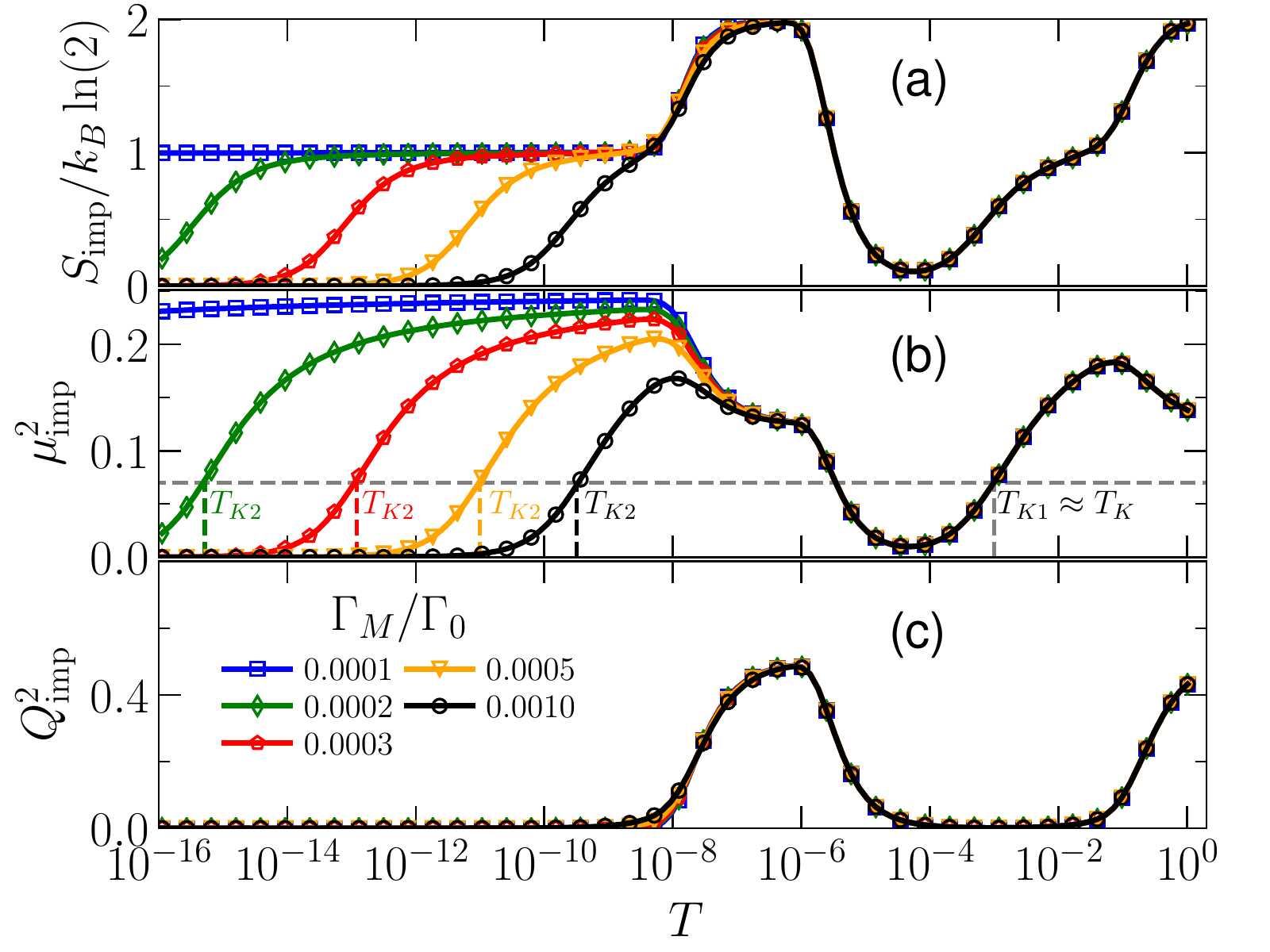}}
	\caption{Impurity contribution to (a) Entropy $S_{\rm imp}$, 
	(b) magnetic moment  $\mu^2_{\rm imp}$, and (c) charge 
	fluctuation $Q_{\rm imp}^2$, as a function of temperature for 
	$10^{-4} < \nicefrac{\Gamma_{\rm M}}{\Gamma_0} < 10^{-3}$ and $\Delta = 10^{-5}$. 
	Note the appearance of a second SC fixed point (for all $\nicefrac{\Gamma_{\rm M}}{\Gamma_0} \geq 0.0002$) 
	at lower temperatures, which can be identified by an increase in charge 
	fluctuation at around $T \approx 10^{-5}$ [panel (c)], 
	followed by an LM regime, followed by an impurity-band singlet formation 
	[panel (b)] at the second SC fixed point, with lowering onset temperature, 
	as $\Gamma_{\rm M}$ decreases. To facilitate the discussion, 
	the estimated values for $T_{K1}$ and $T_{K2}$ (obtained through Wilson's criterion) 
	are indicated in panel (b). See text for details.} 
\label{Fig_thermo_1}
\end{figure}

Let us now turn our attention to the full system, which includes the metallic contact. 
In particular, we are interested in studying what happens to the system for temperatures 
below $T_K$, where, again, $T_K$ is the Kondo temperature for $\Delta=0$ and $V_M=0$. 
To do this, we fix $\Gamma_0=0.05$ and $\Delta=10^{-5}$, 
and vary $\Gamma_{\rm M}$. Note that, as can be checked from the NRG curve in 
Fig.~\ref{PMS} (red squares), for these parameter values we have that $\Gamma_0 > \Gamma_{c}$. 
Our results now rely just on NRG calculations, since 
PMS breaks down before $\tilde D<\Delta$, as shown in the previous section. 
We will see that an intriguing `revival' of an effective Anderson Hamiltonian is observed 
as the temperature tends to zero. This assertion will become clear after we analyze 
the impurity thermodynamic properties, where it will become evident the appearance of 
the two Kondo temperatures mentioned in Sec.~\ref{sec:intro}, $T_{K1}$ and $T_{K2}$, 
with $T_{K1} \gg T_{K2}$ [see Fig.~\ref{Fig_thermo_1}(b)]. In addition, it should be noted that,  
as expected [and indicated in Fig.~\ref{Fig_thermo_1}(b)], the higher Kondo temperature $T_{K1}$, 
obtained for finite $\Gamma_M$ and $\Delta$, has approximately the same value as the Kondo temperature 
$T_K$, corresponding to the $\Gamma_M = \Delta = 0$ case, as long as $\Gamma_M$ and $\Delta$ are $\ll T_K$. 

Figure~\ref{Fig_thermo_1} shows the impurity contribution to the entropy, 
$S_{\rm imp}$ [Fig.~\ref{Fig_thermo_1}(a)], magnetic moment, $\mu^2_{\rm imp}$ [\ref{Fig_thermo_1}(b)], 
as well as the charge fluctuations, $Q_{\rm imp}^2$ [\ref{Fig_thermo_1}(c)], as a 
function of temperature for five different values of $\Gamma_{\rm M}$ in the interval 
$10^{-4} \leq \nicefrac{\Gamma_{\rm M}}{\Gamma_0} \leq 10^{-3}$. 
We first note that, for temperatures in the interval $10^0 > T \gtrsim 10^{-5} = \Delta$, 
all impurity thermodynamic properties are independent of $\Gamma_{\rm M}$, and 
the results display the traditional SIAM behavior, in which 
the system crosses over from the FO to the LM to an SC fixed point, as the temperature 
decreases. These three fixed points are marked, respectively, by entropy values 
$\nicefrac{S_{\rm imp}}{k_B} \sim \ln 4$, $\sim \ln 2$, and $\sim 0$, as seen in Fig.~\ref{Fig_thermo_1}(a). 
This is accompanied by an enhancement of the magnetic moment $\mu_{\rm imp}^2$, at the LM 
fixed  point, followed by its complete suppression in the SC fixed point, as shown in
Fig.~\ref{Fig_thermo_1}(b). Finally, notice also the strong suppression of the impurity 
charge fluctuations $Q_{\rm imp}^2$ (at the LM and SC points) [Fig.~\ref{Fig_thermo_1}(c)]. 
Interestingly, as mentioned above, all these features are independent of the $\Gamma_{\rm M}$ value. This can 
be easily concluded from the superposition of all the curves in all panels in 
Fig.~\ref{Fig_thermo_1} in the temperature interval $10^0 > T\gtrsim 10^{-5}$. 
This behavior may be associated to the fact that the largest $\Gamma_{\rm M}$ used 
in the results shown in Fig.~\ref{Fig_thermo_1} (given by $10^{-3}\Gamma_0=5 \times 10^{-5}$) 
was still much smaller than $T_{K}\approx 10^{-3}$.

It is well-known that the thermodynamic properties presented above (for the temperature interval 
$10^0 > T\gtrsim 10^{-5}$) are characteristic of 
the SIAM~\cite{Hewson-Kondo}. However, for a \emph{traditional} SIAM, the values of the 
thermodynamic quantities, for $T \ll T_K$, i.e., well into the SC regime, remain unchanged 
down to $T\rightarrow 0$, as the system would have already reached the stable SC fixed point 
and would stay there. Remarkably, in the present case, when $T$ approaches $\Delta=10^{-5}$ 
(from above), the system deviates from this standard behavior, as it can be easily seen in 
Fig.~\ref{Fig_thermo_1}, since all thermodynamic properties have additional structures for 
$T<\Delta$. Indeed, when $T\rightarrow \Delta$, the system flows to a \emph{second} free orbital 
(SFO) fixed point, marked by an increase of $S_{\rm imp}$, $\mu^{2}_{\rm imp}$, and $Q_{imp}^2$, 
to values that go back to their high temperature ($T=D$) values. Further decrease of $T$ shows 
that the system crosses over fixed points that have very similar properties to the ones 
crossed in the temperature interval $10^0 > T \gtrsim 10^{-5}$. The similarity between 
the low and high temperature fixed points indicates that, for $T < \Delta$, the system 
seems to be governed by an \emph{effective} SIAM with renormalized parameters and a 
much lower Kondo temperature. 
Note that the extent of the plateaus in the entropy (at $k_B\ln 2$) and in the magnetic moment 
(at $\approx \nicefrac{1}{4}$), which mark how long the system stays close to the LM 
fixed point, depend strongly on $\Gamma_{\rm M}$, 
showing that the Kondo temperature for the `reentrant' effective SIAM, denoted as $T_{K2}$, 
depends strongly on $\Gamma_{\rm M}$. To highlight that, in Fig.~\ref{Fig_thermo_1}(b) 
we use Wilson's criterion to determine the characteristic Kondo temperatures 
$T_{K1}$ and $T_{K2}$, which can be extracted from the intersection of the gray dashed line (corresponding to  
$\mu^{2}_{\rm \imp}=0.07$) with the $\mu^{2}_{\rm \imp}$ curves for different $\Gamma_{\rm M}$ values. 
The higher Kondo temperature, $T_{K1}$, indicated on the right side of panel (b), which is clearly 
independent of $\Gamma_M$, and similar to the $\Delta=\Gamma_{\rm M}=0$ Kondo temperature $T_K$, 
is accompanied by a $\Gamma_{\rm M}$-dependent $T_{K2}$ Kondo temperature, much lower than $T_{K1}$ 
and associated to a stable SC fixed point. Thus, the thermodynamic quantities 
($S_{\rm imp}$, $\mu^2_{\rm imp}$, and $Q^2_{\rm imp}$) exhibit a 
behavior compatible with an NRG flow through a low temperature \emph{second stage} effective SIAM, as will 
be explicitly shown next.  

\begin{figure}[ht!]
\centering
\subfigure{\includegraphics[clip,width=3.4in]{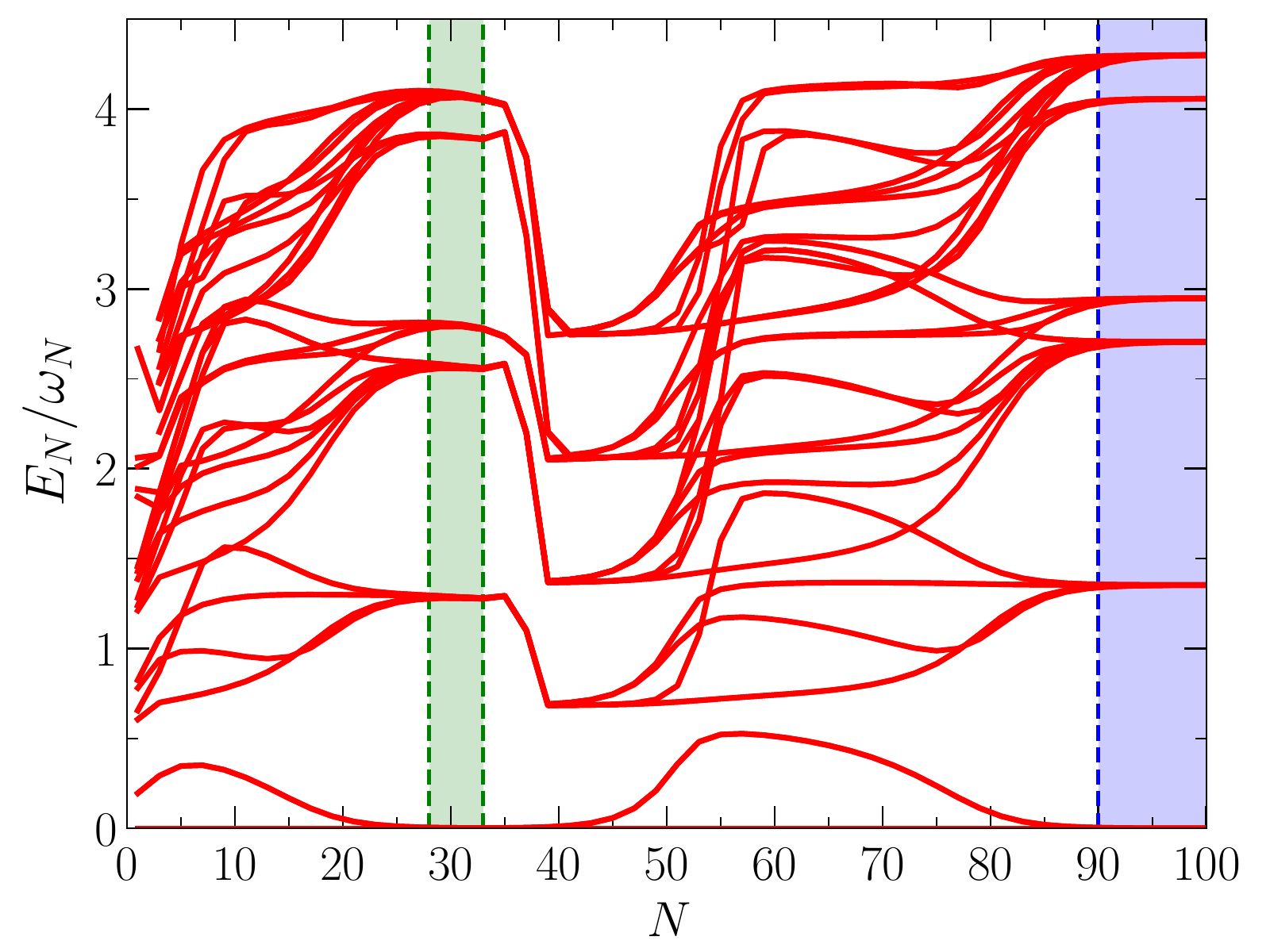}}
\caption{Energy spectrum vs NRG iteration step $N$ 
	(odd values only) obtained for the lowest energy 
	levels. Note the 
	fixed points in the traditional Anderson model seen in 
	the iterations ranging from $N \approx 5$ to $N\approx 35$, which 
	are traversed again at higher N-values ($N \gtrsim 41$), 
	showing the reentrance of the Anderson model behavior at low 
	energies. The model parameters used here were  
$\Gamma_0=0.05$, $\Gamma_{\rm M}/\Gamma_0=5\times 10^{-4}$.} 
\label{Fig_FLOW}
\end{figure} 

Indeed, this interesting (and unusual) behavior can be clearly captured by the energy 
flow diagram obtained from NRG, as shown in Fig.~\ref{Fig_FLOW}, which displays the 
energy spectrum as function of the NRG iteration step $N$ (for odd values). 
As described in Ref.~\cite{Krishna-murthy1980}, the occurrence of a fixed point in the 
iterative NRG procedure can be determined by looking for a set of many-particle energy levels 
that repeat themselves in a sequence of odd (or even) steps in the NRG diagonalization procedure. 
Figure \ref{Fig_FLOW} shows that the traditional 
SIAM fixed points are observed in the range of iterations from $N \approx 5$ to $N\approx 35$, 
while the second stage SIAM fixed points are traversed again at higher $N$-values  
($N \gtrsim 41$). For the sake of clarity, we added a green-shaded vertical stripe to 
highlight the (unstable) SC fixed point and a blue-shaded one to highlight the second (stable) SC fixed point. 
The parameters used were $\Gamma_0=0.05$, $\Gamma_{\rm M}/\Gamma_0=5\times 10^{-4}$, the same 
as for the inverted triangle curves in Fig.~\ref{Fig_thermo_1}. 
\begin{figure}[ht!]
\centering
\subfigure{\includegraphics[clip,width=3.4in]{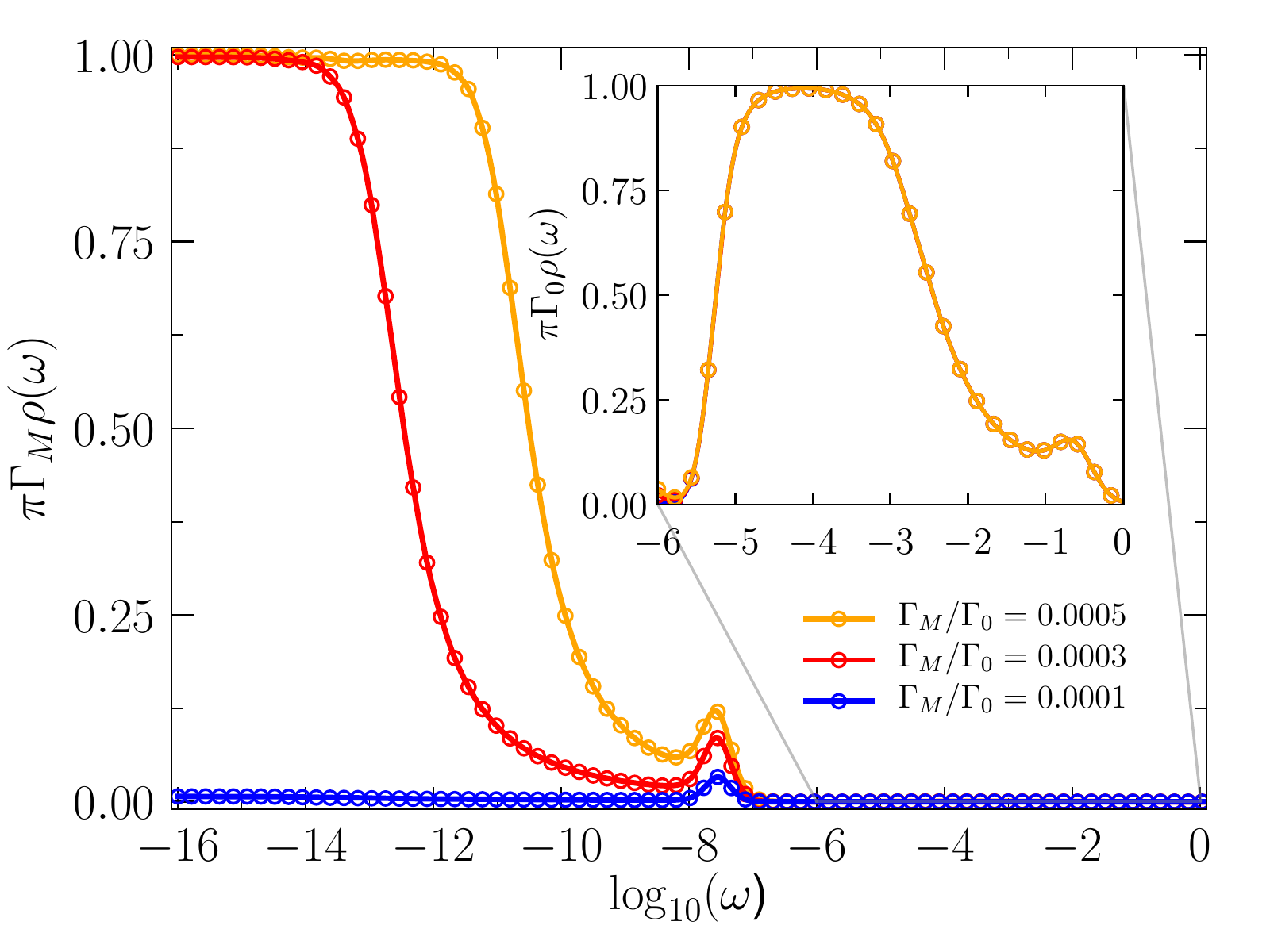}}
\caption{Impurity LDOS as a function of energy for $\Gamma_0=0.05$ and three  
	values of $\Gamma_{\rm M}$. The inset shows a zoom-in of the region where the first Kondo 
	regime occurs. The LDOS $\rho(\omega)$, in the main panel and in the inset, is multiplied by $\pi \Gamma_{\rm M}$ 
	and $\pi\Gamma_0$, respectively, so as to show that both Kondo regimes obey the Friedel sum rule.} 
\label{Fig_DOS}
\end{figure} 

Further insight onto the two SC fixed points can be gained from the analysis of the impurity's 
LDOS, given by   
\begin{eqnarray}\label{LDOS}
	\rho(\omega)=-\frac{1}{\pi}{\rm Im}[\greenfunc{d_\sigma}{d^\dagger_{\sigma}}_{\omega}],
\end{eqnarray}
where $\greenfunc{d_\sigma}{d^\dagger_{\sigma}}_\omega$ is the retarded local Green's function 
in the energy domain, within Zubarev's notation~\cite{Zubarev_1960}. 
We first analyze the impurity LDOS at low energies ($\omega < 10^{-7}$) in the main panel of 
Fig.~\ref{Fig_DOS}, which shows $\pi\Gamma_{\rm M}\rho(\omega)$ 
as a function of $\log_{10}\omega$ for three values of $\Gamma_{\rm M}$. For $\Gamma_{\rm M}=3\times 10^{-4}\Gamma_0$ 
and $\Gamma_{\rm M}=5\times 10^{-4}\Gamma_0$ (red and orange curves, respectively), we see Kondo 
peaks that nicely obey the Friedel sum rule. Notice that, to accomplish this, we are multiplying 
$\rho(\omega)$ by $\pi \Gamma_{\rm M}$, the impurity-coupling to the metallic lead. This shows that the reentrant Kondo 
state, as expected, involves electrons from the metallic DOS. However, contrary to what happens for the two larger values 
of $\Gamma_{\rm M}$, for $\Gamma_{\rm M}=10^{-4}\Gamma_0$ (blue curve), there is no Kondo 
peak at low energies (at least down to $\omega=10^{-16}$). This is in agreement with 
the thermodynamic properties for the corresponding (blue) curves in Fig.~\ref{Fig_thermo_1}, 
which show no indication of the occurrence of a reentrant Kondo effect. In addition, 
the width of the two Kondo peaks in the main panel of Fig.~\ref{Fig_DOS}, for 
$\Gamma_{\rm M}=3\times 10^{-4}\Gamma_0$ and $\Gamma_{\rm M}=5\times 10^{-4}\Gamma_0$ 
are in accordance with the estimated values for $T_K$ using Wilson's criterion in panel (b) of 
Fig.~\ref{Fig_thermo_1}.
Finally, it is interesting to notice that the small peaks observed slightly above $\omega=3\times 10^{-8}$ 
correspond to the upper Hubbard peak, which is located at $\nicefrac{U_{\rm eff}}{2}$, where the renormalized 
Coulomb repulsion $U_{\rm eff}$ is associated to the effective reentrant SIAM (see more details below).

We now proceed to an analysis of the LDOS at higher values of $\omega$. The inset in 
Fig.~\ref{Fig_DOS} shows a zoom of the $\omega \in [10^{-6},1]$ energy window. 
Note that, in accordance with the thermodynamic quantities analyzed 
in Fig.~\ref{Fig_thermo_1}, all three curves collapse onto each other. 
In addition, as was the case at lower energies (main panel), if one 
multiplies $\rho(\omega)$ by $\pi\Gamma_0$ (as done in the inset), 
the results obey the Friedel sum rule, indicating that, for the first SC fixed point, 
the many-body state is formed between the impurity and the electrons from the semiconducting 
DOS. The interpretation here is immediate: the higher peak corresponds to the first ($T_{K1}$) 
Kondo effect, while the smaller peak  above $\omega=10^{-1}$ 
corresponds to the upper Hubbard  peak, located at $\nicefrac{U}{2}$. 

The LDOS results just presented in Fig.~\ref{Fig_DOS} provided access to the
numerical value of $U_{\rm eff}$ (the small peak in the main panel). Since it, together with $\Gamma_{\rm S}$,
$\Gamma_{\rm M}$, and $U$, characterizes the thermodynamic properties
shown in Fig.~\ref{Fig_thermo_1}, we will, in what follows, correlate (and summarize) 
the results presented in Fig.~\ref{Fig_thermo_1} with those presented in Fig.~\ref{Fig_DOS}. In Fig.~\ref{Fig_thermo_1}, 
one can clearly see that, as the temperature decreases below the first Kondo 
temperature $T_{K1} \approx 10^{-3}$, the system enters the SFO fixed point (for 
$T \approx \Delta=10^{-5}$), where the coupling between the impurity and the conduction electrons drops 
from $\Gamma_0$ to $\Gamma_M$, in which case we have that 
$T \gtrsim \Gamma_{\rm M}$, and $T \gg U_{\rm eff}~\approx 3\times 10^{-8}$ (see Fig.~\ref{Fig_DOS}). 
As the temperature decreases further, the system then enters the second LM fixed point for 
$T \lesssim U_{\rm eff}~\approx 3\times 10^{-8}$ (compare Figs.~\ref{Fig_thermo_1} and~\ref{Fig_DOS}). 
Finally, when $T$ goes below the second Kondo temperature ($T_{K2}$, whose value depends strongly on 
$\Gamma_{\rm M}$, see Figs.~\ref{Fig_thermo_1}(b) and main panel of \ref{Fig_DOS}) the system reaches the stable SC fixed point. 

The existence of this very small $U_{\rm eff}$ can be inferred from the PMS
analysis of the Anderson model, as discussed by Jefferson~\cite{Jefferson_1977} 
and Haldane~\cite{Haldane1978} for metallic conduction bands, 
and, later on, extended to more 
general spectra in Refs.~\cite{Cheng2013} and~\cite{Kevin2017} 
(see, for instance, Eq.~(27) of Ref.~\cite{Kevin2017}). 
Although these analyses are limited by their perturbative character, they suggest that 
the renormalized Coulomb repulsion indeed decreases along the RG flow. 

\begin{figure}[ht!]
\centering
\subfigure{\includegraphics[clip,width=3.4in]{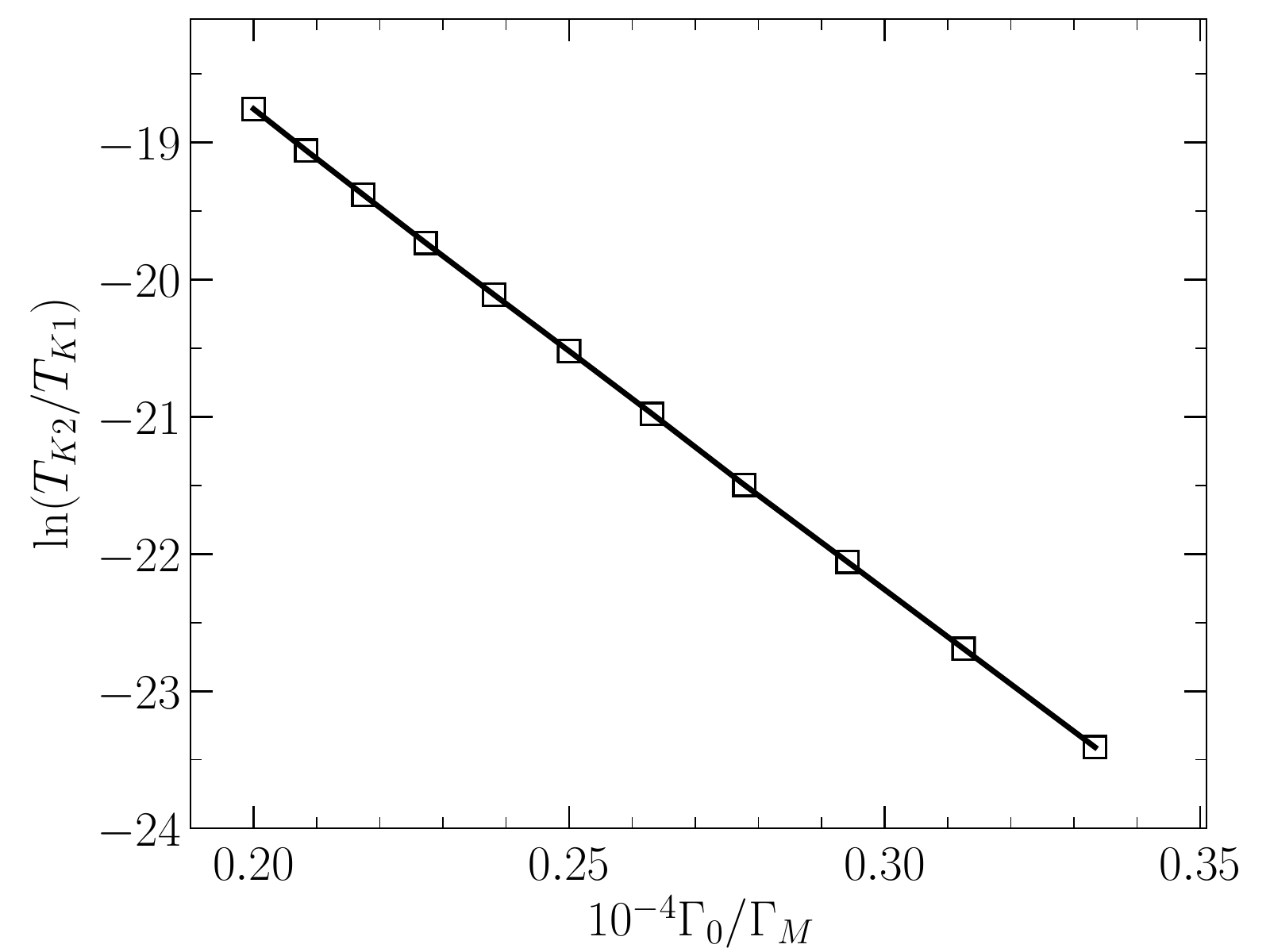}}
\caption{$\ln(T_{K2}/T_{K1})$ vs $10^{-4} \times \Gamma_0/\Gamma_{\rm M}$, for $\Gamma_0=0.05$. 
	From the linear behavior of the curve, the data could be fitted to an expression like 
	$T_{K2}=A_0 e^{-A_1/\Gamma_{\rm M}}$. } 
\label{Fig_TK2xGamma}
\end{figure}

Since the width of the Kondo peak at half-height is a good estimate of the 
Kondo temperature, calculations for various values of $\Gamma_{\rm M}$, at fixed $\Gamma_0$, like the 
ones done in Fig.~\ref{Fig_DOS}, provide the dependence of the Kondo temperature of the reentrant Kondo screening, $T_{K2}$, 
on $\Gamma_{\rm M}$. These results are shown in Fig.~\ref{Fig_TK2xGamma}, where we 
plot $\log(T_{K2}/T_{K1})$ as a function of $\Gamma_0/\Gamma_{\rm M}$ (for $\Gamma_0=0.05$, in units of $10^{-4}$). 
The remarkable linear behavior of the curve suggests a fitting of the NRG results to an expression like 
$T_{K2}=A_0 e^{-A_1/\Gamma_{\rm M}}$, where both $A_0$ and $A_1$ are positive and $A_0 \propto T_{K1}$. 
This expression indicates that $T_{K2}$ decreases exponentially with a decreasing $\Gamma_{\rm M}$. 
The parameters $A_0$ and $A_1$ contain the intricate information about the reentrant effective SIAM. 

\begin{figure}[h!]
\centering
\subfigure{\includegraphics[clip,width=3.4in]{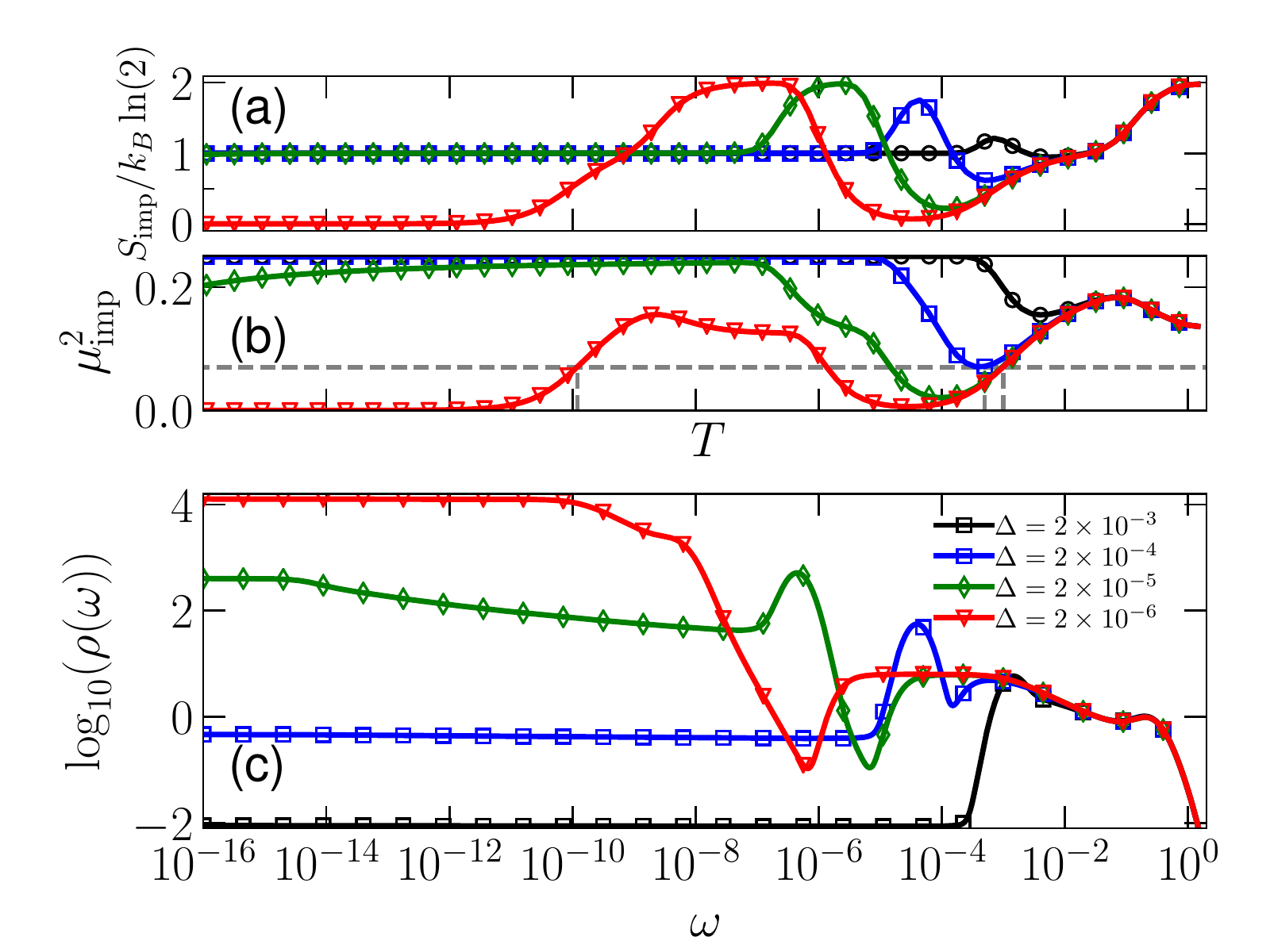}}
\caption{(a) Impurity entropy $S_{\rm imp}$ and (b) magnetic moment $\mu_{\rm imp}^{2}$, 
	as a function of $T$, and (c) $\log_{10}[\rho(\omega)]$ vs energy, for $\Gamma_0=0.05$, 
	$\Gamma_{\rm M}=5\times 10^{-4}$, and four $\Delta$ values ($2.0 \times 10^{-6} \leq \Delta \leq 2 \times 10^{-3}$). 
	The horizontal gray dashed line in panel (b) represents $\mu^{2}_{imp}=0.07$, and 
	from its intersection with the $\mu^{2}_{imp}$ curves we obtain $T_{K1}$ and $T_{K2}$ 
	for each $\Delta$ value. In panel (c), we have chosen to show $\log_{10}[\rho(\omega)]$ 
	to visualize all the peaks, as their height differ by several orders of magnitude. 
	Note that the horizontal axis scale (not shown) in panels (a) and (b) is the same 
	as in panel (c). 
	}
\label{Fig_Thermo_gap}
\end{figure} 

Before closing this section, in Fig.~\ref{Fig_Thermo_gap} we show how both Kondo screenings 
change, in respect to the gap $\Delta$ in the semiconducting lead. Panels (a), (b), and (c), in Fig.~\ref{Fig_Thermo_gap},  
show the impurity entropy $S_{\rm \imp}$, magnetic moment $\mu^{2}_{\rm \imp}$, and 
LDOS $\rho(\omega)$, respectively, for four different $\Delta$ values 
($2.0 \times 10^{-6} \leq \Delta \leq 2 \times 10^{-3}$). The calculations were done 
for $\Gamma_0 = 0.05$ and $\Gamma_{\rm M} = 5 \times 10^{-4}$, which is an order of magnitude above the 
largest $\Gamma_{\rm M}$ value used in Fig.~\ref{Fig_thermo_1}. 
Notice that, in Fig.~\ref{Fig_Thermo_gap}(b) (as done also in Fig.~\ref{Fig_thermo_1}), the characteristic Kondo temperatures 
$T_{K1}$ and $T_{K2}$, for each value of $\Delta$, can be extracted from the 
intersection of the gray dashed line (corresponding to  
$\mu^{2}_{\rm \imp}=0.07$) with the $\mu^{2}_{\rm \imp}$ curves. It is straightforward to note that, 
for the smallest value of $\Delta$ analyzed [$\Delta=2.0 \times 10^{-6}$ 
(red curve)], $S_{\rm imp}$ and $\mu_{\rm imp}^{2}$ are strongly suppressed in the temperature 
interval $10^{-5}\lesssim T \lesssim 10^{-4}$ and vanish as $T \rightarrow 0$ 
(below $T \approx 10^{-11}$), clearly showing the existence 
of two Kondo screening regimes, the first with $T_{K1} \approx 10^{-3}$ and the second with $T_{K2} \approx 10^{-10}$ 
[as indicated in panel (b)]. The impurity LDOS [panel (c)] for the same value of $\Delta = 2.0 \times 10^{-6}$ (red curve) 
exhibits, accordingly, two (not normalized) Kondo peaks, with respective heights  
$\nicefrac{1}{\pi\Gamma_0}$ and $\nicefrac{1}{\pi\Gamma_{\rm M}}$, for the first and second 
Kondo regimes, respectively. However, for the larger $\Delta$ values shown in Fig.~\ref{Fig_Thermo_gap}, 
we note that the first Kondo regime is progressively suppressed. This occurs because, as $\Delta$ 
increases, $\Gamma_c$ also increases, eventually becoming larger than $0.05$, the $\Gamma_0$ value 
used in the calculations (see NRG results (red squares) in Fig.~\ref{PMS}). Fig.~\ref{Fig_Thermo_gap}(a) 
shows the details of how this behavior evolves. First, it is important to remark that, as shown 
in Fig.~\ref{Fig_thermo_1}, the end of the first Kondo screening occurs 
for $T \approx \Delta$. Second, as can be seen in Fig.~\ref{Fig_Thermo_gap}, 
panels (a) and (b), the temperature at which the transition 
from the LM to the SC fixed point starts, for the first Kondo stage, does not depend on $\Delta$. Thus, 
as $\Delta$ increases, the flow from LM to SC is cut short and the $T\rightarrow0$ physics 
is that of the first LM fixed point (i.e., $S_{\imp}=k_B\ln2$ and $\mu^{2}_{imp}=\nicefrac{1}{4}$). 
In other words, 
the first SC fixed point is squeezed out of existence by the increase in $\Delta$ and 
the system gets stuck in the first LM fixed point. The two $\Delta = 2.0 \times 10^{-6}$ 
Kondo peaks shown in the LDOS (panel (c), red curve), in turn, are progressively suppressed 
as $\Delta$ increases (see the green, blue, and black curves in Fig.\ref{Fig_Thermo_gap}(c)), 
confirming the destruction of both Kondo screening regimes. Thus, the first LM fixed point becomes the low temperature stable fixed point.
 
The results shown so far are quite general and may be applicable to a variety of gapped systems 
to which a magnetic impurity can be coupled to. Examples encompass narrow-gap 
semiconductors~\cite{Massidda1990}, synthesized polymers~\cite{Heeger1988}, as well as  
modern gap-engineered materials~\cite{Borghardt2017}. In the following, we shall 
discuss how the reentrant SIAM behavior emerges in an 
AGNR in which a Rashba spin-orbit coupling (and thus a gap) is induced externally~\cite{Lenz2013}.

\section{Reentrant Kondo Effect in Armchair Graphene Nanoribbon}
\label{sec:AGNR}
In this section, we discuss a plausible experimental setup consisting of a 
magnetic impurity coupled to an AGNR, subjected to a tunable spin-orbit coupling, in which the phenomena presented  
in Sec.~\ref{sec:reent} may be experimentally observed.

\begin{figure}[th!]
\begin{center}
\includegraphics[scale=0.5]{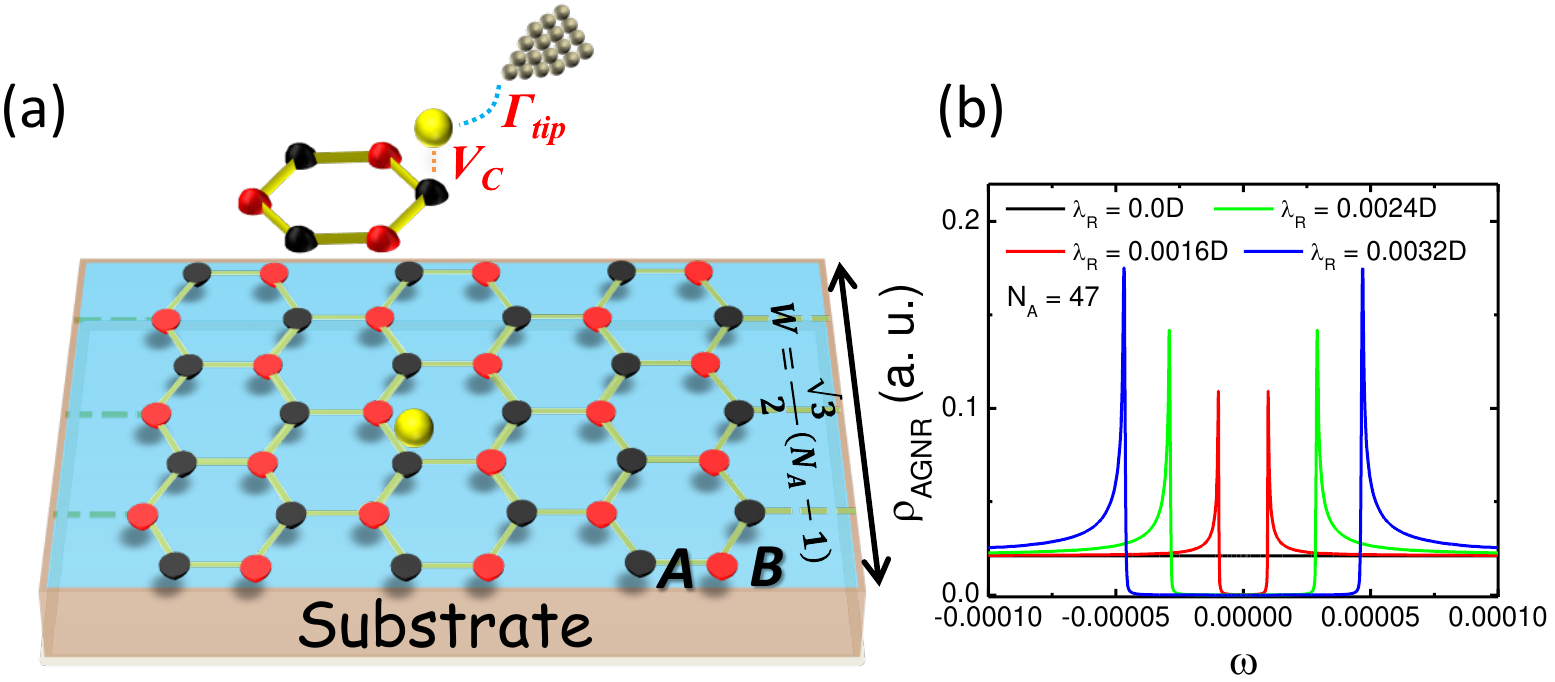}
\caption{(a) Schematic representation of an $N_{A}$-AGNR deposited on a substrate, 
	with a magnetic impurity (yellow) deposited in a top-site configuration (right above a nanoribbon 
	carbon atom (black), and strongly coupled to it, with hopping amplitude $V_C$). 
	Right on top of the magnetic impurity adatom (as shown 
	in the inset) is located a weakly coupled \emph{metallic} STM tip, with a coupling 
	strength $\Gamma_{\tip}$. (b) 
	DOS for a 47-AGNR close to the Fermi level, without the impurity, as a function of energy $\omega$, for 
	different RSOI strengths $\lambda_R$. $W$ is the width of the AGNR (assuming a nearest-neighbor 
	distance $a_{C-C}=1$), which depends on the number of dimmers, $N_A$, across the nanoribbon. 
	Note that, as $N_A=47=3 \times 16 - 1$, the $\lambda_R = 0.0$ DOS (black curve) is metallic, 
	while a finite $\lambda_R$ opens a gap in the spectra.}
\label{fig1}
\end{center}
\end{figure}

It has been shown recently by Lenz et al.~\cite{Lenz2013} that, under the 
influence of Rashba spin-orbit interaction (RSOI), due to an external electric field, or induced by a 
substrate, AGNRs exhibit a tunable band gap at the Fermi level~\cite{Dimmers}. 
In the following, we will consider a magnetic impurity coupled to such a gapped 
AGNR and weakly coupled to an STM tip (see Fig.~\ref{fig1}). 
By employing a tight-binding model, combined with NRG calculations, we  show that 
this setup is very convenient to investigate the reentrant Kondo effect discussed in Sec.~\ref{sec:reent}. 

It is important to notice that, as already mentioned above, an AGNR may be 
metallic (when the number of dimers $N_A$ across its width $W$ is such that $N_A = 3M + 1$, 
where M is an integer), or semiconducting (for other values of $N_A$). The use 
of an \emph{intrinsic} semiconducting AGNRs for the purpose of testing the reentrant Kondo effect would 
be problematic for two reasons: first, the typical gap values $\Delta$ that one obtains are in general 
\emph{large}, and second, they are hard to tune. The proposal of using RSOI to produce a small and tunable gap 
$\Delta$ in a metallic AGNR, as illustrated in Fig.~\ref{fig1}(b), sidesteps both problems at once. 

Our proposed setup is schematically shown in Fig.~\ref{fig1}(a). The system, comprised of a single magnetic 
impurity coupled to an AGNR, is modeled by the standard SIAM-like Hamiltonian~\cite{Anderson1961}, given by
\begin{align}
	\label{H1}
	H \!=\! H_{\rm AGNR} + H_{\rm imp}+H_{\rm tip} + H_{\rm AGNR-imp} +H_{\rm
		imp-tip},
\end{align}
where the first term describes the AGNR, which is modeled by a tight-binding 
Hamiltonian in real space, given by 
\begin{eqnarray}
	H_{\rm AGNR}&=&\sum_{i\sigma}(\varepsilon_0 - \mu)c^\dag_{i\sigma}c_{i\sigma} + \sum_{\langle	i,j\rangle,\sigma\sigma^{\prime}}\left[t_{ij}\delta_{\sigma\sigma^{\prime}} + \right.\\\nonumber &&\left.i\lambda_{R}\hat{\bf z}\cdot ({\bf s}\times {{\bm \delta}}_{ij})\right]c_{i\sigma}^{\dagger} c_{j\sigma^{\prime}},
	\label{hagnr}
\end{eqnarray}
where $c_{i\sigma}^{\dagger}$ ($c_{i\sigma}$) creates (annihilates) an electron with 
energy $\varepsilon_0$ and  spin $\sigma$ on the $i$-th site of the AGNR, 
and $\mu$ is the chemical potential, which can be externally tuned by a back gate. 
The second term is the nearest-neighbor $\pi$-band tight-binding Hamiltonian, 
where $t_{ij}=t_{0}$ is the hopping between nearest-neighbor sites~\cite{PhysRevB.80.045401}, 
with $t_{0}\approx 2.7$eV~\cite{RevModPhys.81.109}. The third term models the induced RSOI, with 
parameter $\lambda_{R}$ proportional to the electric field applied perpendicular to 
the $x$-$y$ plane of the nanoribbon~\cite{Kane2005,PhysRevB.79.165442}, 
${\bf s}=(s_x,s_y,s_z)$ represents a vector of Pauli spin matrices and ${\bm \delta}_{ij}$ 
are the vectors connecting nearest neighbor sites. The second term of Eq.~\eqref{H1} 
describes the single level Anderson impurity (given by Eq.\eqref{H_imp}, in Sec.~\ref{sec:model}), 
while the third term describes the STM tip, which is modeled by the Hamiltonian 
$H_{\rm M}$ in Eq.~\eqref{Anderson_Band}. The fourth 
term in Eq.~\eqref{H1}, which couples the impurity to the AGNR, is given by 
\begin{align}
	\label{H2} H_{\rm AGNR-imp} = \sum_{j,\sigma}V_{j\sigma}\left(c_{j\sigma}^{\dagger}d_{\sigma}+{\rm H.c.} \right),
\end{align}
where the most general situation is that in which the index $j$ runs over a number of sites in the AGNR 
that are closest to the impurity. In Fig.~\ref{fig1}(a), we depict the situation where the impurity couples to just one site. 
Finally, the last term in Eq.~\eqref{H1}, which couples the impurity to the STM tip, reads as 
\begin{eqnarray}
	\label{H3}
	H_{\rm imp-tip}=\sum_{{\bf k}\sigma}\left(V_{\bf k}c^\dagger_{\bf k
		\sigma}d_{\sigma}+{\rm H.c.} \right).
\end{eqnarray}
In Eq.~\eqref{H2}, if we consider the situation depicted in Fig.~\ref{fig1}, 
where the impurity couples to a single carbon atom in the ribbon, 
then, assuming that the RSOI has no effect over this coupling 
(thus, the coupling is spin independent), we can set $V_{j\sigma}\equiv V_{C}$. 
Furthermore, assuming a constant density of states at the metallic tip, 
$\rho_{\rm tip}$, we may write the tip-impurity hybridization function as 
$\Gamma_{\tip} =\pi V_{\rm tip}^2\rho_{\rm tip}$, where $V_{\rm tip}$ is the 
hopping parameter between the impurity and the STM tip. Thus, $\Gamma_{\rm tip} \equiv \Gamma_{\rm M}$, 
as defined in Sec.~\ref{sec:model}. Therefore, from now on, to facilitate the comparison 
with the results in Sec.~\ref{sec:model}, we will denote the QI-STM coupling by $\Gamma_{\rm M}$ 
(instead of $\Gamma_{\tip}$) to present all the forthcoming results. 

To perform the NRG calculations to tackle the Kondo effect in this system, we need to 
calculate the hybridization function $\Gamma_0(\omega)$~\cite{Krishna-murthy1980,Bulla2008}. 
To do that, we have implemented a recursive Green's function approach~\cite{Nardelli1999,Sancho1984} 
for the non-interacting case, i.e., $U=0$. Having the local Green's function at hand~\cite{PhysRevB.97.115444}, we can 
obtain the self-energy matrix for the impurity, $[{\bm \Gamma}_{\rm AGNR+tip}]_{\sigma\sigma^{\prime}}
(\omega)={\rm Im}[{\bf G}^{-1}_{\rm C+tip}(\omega)]_{\sigma \sigma^{\prime}}$, where 
${\bf G}_{\rm C+tip}$ is the AGNR+tip~\cite{AGNR+tip} non-interacting, local (at the impurity site), Green's function matrix. 
We assume the magnetic impurity placed at a top-site configuration~\cite{topsite}, as depicted in Fig.~\ref{fig1}, 
in which case the system is still bipartite and the particle-hole symmetry of the whole system is 
preserved~\cite{PhysRevLett.92.216401}. This is important, as it allows for a direct comparison of the results 
in this section with those in Sec.~\ref{sec:model}. Finally, note that, as the RSOI does not 
break time-reversal symmetry,  we have that the ${\bm \Gamma}_{\rm AGNR+tip}$ matrix is diagonal, 
thus $[{\bm \Gamma}_{\rm AGNR+tip}]_{\uparrow \uparrow} = [{\bm \Gamma}_{\rm AGNR+tip}]_{\downarrow \downarrow} 
\equiv \Gamma_0$~\cite{AGNR+tip}. 

For concreteness, we consider a metallic AGNR, of width $W=\sqrt{3}\left(N_A-1\right)/2$, 
where $N_{A}$ is the number of dimmers along the transverse direction [see Fig.~\ref{fig1}(a) for details].
Moreover, we have chosen the carbon-carbon hopping amplitude $t\approx1/3.1$, so that the 
half bandwidth is $D=1$, thus consistent with Sec.~\ref{sec:model}, where the half bandwidth was taken as the 
energy unit. 
Figure \ref{fig1}(b) shows the DOS $\rho_{\rm AGNR}(\omega)$ for a 47-AGNR, close to the 
Fermi level, for a pristine nanoribbon, i.e., without any impurity coupled to its surface, 
for different values of RSOI. We clearly see that in the absence of RSOI ($\lambda_R=0$) 
our AGNR exhibits a gapless DOS as shown by the black line in Fig.~\ref{fig1}(b). 
However, a finite $\lambda_R$ induces a gap $\Delta$ around the Fermi level as shown by the red 
($\lambda_R=1.6 \times 10^{-3}$), green ($\lambda_R=2.4 \times 10^{-3}$), 
and blue ($\lambda_R=3.2 \times 10^{-3}$) curves in Fig.~\ref{fig1}(b), for progressively larger values of $\lambda_R$. 
Thus, the AGNR with finite RSOI simulates 
the semiconducting band coupled to the impurity, while the STM tip plays the role of the metallic band  
defined in Sec.~\ref{sec:model}, introducing a small but finite broadening of the 
impurity level, $\Gamma_{\M}$, inside the gap. It is worthwhile to remark that: (i) the 
RSOI-induced gap $\Delta$ has a particular dependence for narrow AGNRs as a function of $\lambda_{R}$, 
specially for large values of $\lambda_{R}$ \cite{Lenz2013}. However, $\Delta$ decreases as the width of a 
metallic AGNR increases, such that, in the limit where border effects over the electronic structure vanish, 
the spin degeneracy will be lifted, but with no band gap, 
as expected for \emph{bulk} graphene~\cite{Kane2005,PhysRevB.79.161409}; (ii) $\Delta$ exhibits a small 
oscillation as a function of $\lambda_{R}$~\cite{Lenz2013}. In our calculations, we restrict  
$\lambda_{R}$ to a range within which $\Delta$ increases monotonically with $\lambda_{R}$ (for a fixed width), and, 
importantly, in agreement with experimental RSOI values in graphene~\cite{PhysRevLett.100.107602,
PhysRevLett.108.066804,Marchenko2012}. 

\begin{figure}[b!]
\begin{center}
\includegraphics[clip,width=3.4in]{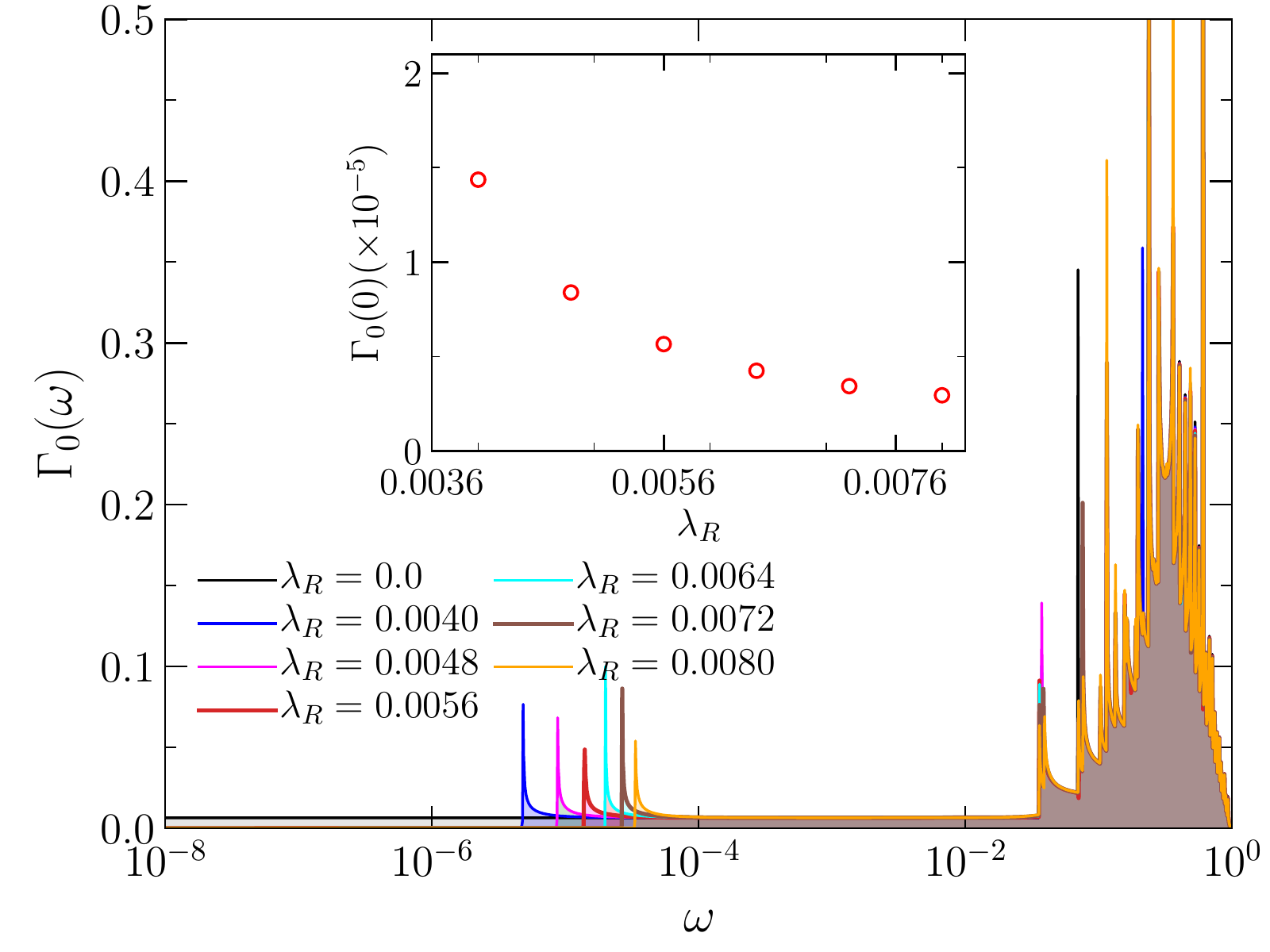}
\caption{Hybridization function $\Gamma_{0}(\omega)$ 
	for vanishing $\lambda_{R}$ (black curve) and in the interval $0.004 \leq \lambda_{R} \leq 0.008$. 
	The range of values of $\lambda_R$ was chosen in order to produce $\Delta$ values 
	monotonically increasing with $\lambda_R$. The 
	inset shows $\Gamma_{0}(0)$ as a function of $\lambda_{R}$. 
	Parameter values are $V_{C}=0.258$ and $\Gamma_{\rm M}=1.0 \times 10^{-6}$.}
\label{delta}
\end{center}
\end{figure}
In what follows, we set $\Gamma_{\rm M} = 1.0\times 10^{-6}$ (thus, fixing $V_{\rm tip}$), 
$V_{C}=0.258$, and $N_{A}=47$ (corresponding to $W\approx$ 5.65nm).
Differently from the case of a zigzag graphene nanoribbon, where the hybridization function 
is strongly dependent on what site (across the ribbon) one chooses to couple the impurity 
to~\cite{PhysRevB.97.115444} (i.e., close or away from the nanoribbon's edge), for an AGNR we have noticed a small quantitative 
difference, as the $\rho_{\rm AGNR}(\omega)$ along the width has a small variation. Therefore, we considered the impurity 
position fixed at a given top-site location~\cite{topsite} for all 
the following calculations. The resulting hybridization function $\Gamma_0(\omega)$, 
for various values of $\lambda_{R}$, is shown in Fig.~\ref{delta}. 
To make the region near the Fermi level (located slightly to the left of the left axis) more visible, 
we plot the energy axis in log-scale, restricted to $\omega > 0$ [by virtue of 
particle-hole symmetry, we have that $\Gamma_0(-\omega)=\Gamma_0(\omega)$]. As expected, for $\lambda_{R}=0.0$ 
the AGNR is metallic, therefore $\Gamma_0(\omega)$ has a constant value ($\approx 0.01$) around 
the Fermi level. In this case, our system behaves quite similarly to a QI coupled to a  
metallic DOS with a flat band. However, for finite $\lambda_R$ we clearly see  
the formation of a small gap $\Delta$, which increases with $\lambda_R$. 
In the inset of Fig.~\ref{delta} we show how 
$\Gamma_0(0)$ evolves with $\lambda_R$. We note that $\Gamma_0(0)$ has a small 
residual and finite value inside the RSOI induced gap, originating from 
the \emph{localized} impurity state contribution, which decreases as $\lambda_R$ 
(or $\Delta$) increases, eventually saturating at $\Gamma_{0}(0)\approx\Gamma_{\rm M}=1.0 \times 10^{-6}$. 
This behavior results from a mixing of spin channels in the conduction band mediated by the 
RSOI, reducing the spin preserving transmission at the Fermi level, as 
when RSOI is switched on the spin-flip mechanism is allowed in the AGNR. 
This band-gap-induced RSOI will show its fingerprints in the impurity thermodynamic 
properties, determining the reentrant SIAM behavior.

\begin{figure}[h!]
\begin{center}
\includegraphics[clip,width=3.4in]{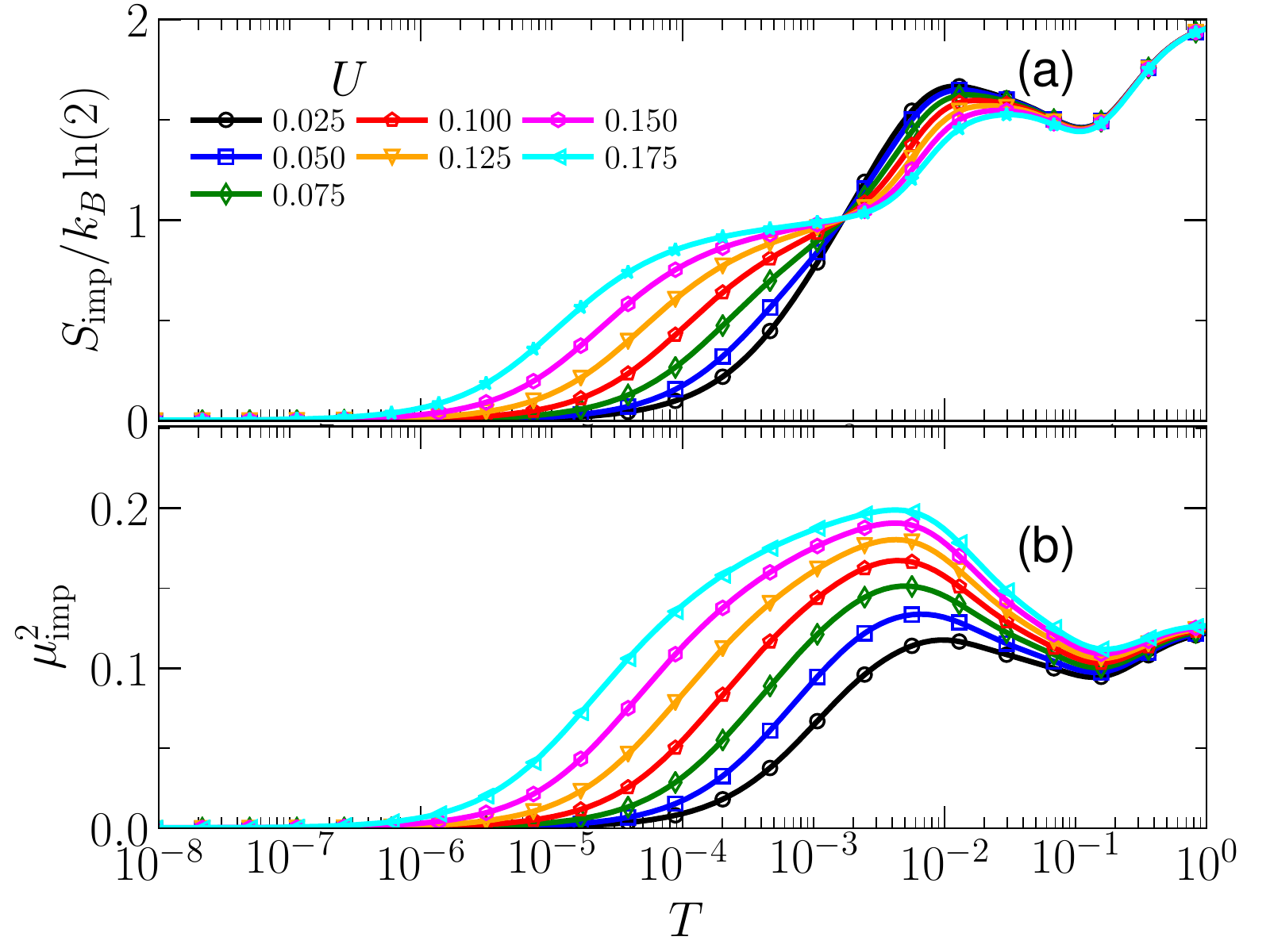}
\caption{(a) Impurity entropy $S_{\rm imp}$ and (b) magnetic moment $\mu_{\rm imp}^{2}$, 
	for a metallic ($\lambda_{R}=0.0$) 47-AGNR,  
	as a function of temperature, for $0.025 \leq U \leq 0.175$, $V_{C}=0.258$, 
	and $\Gamma_{\rm M}=1.0 \times 10^{-6}$.}
\label{fig1-AGNR}
\end{center}
\end{figure}
Before studying how the induced gap affects the Kondo screening in the system, let us first 
analyze the Kondo effect in the absence of RSOI, and then see how it is modified by a finite RSOI. 
In Fig.~\ref{fig1-AGNR}, we show, in panel (a), the impurity entropy contribution, $S_{\rm imp}$, 
and, in panel (b) the magnetic moment, $\mu_{\rm imp}^{2}$, both of them as a function of temperature 
($10^{-8} < T < 1$), for $\lambda_{R}=0.0$ and 
$0.025 \leq U \leq 0.175$. As expected, the characteristic behavior of the SIAM is observed 
as the temperature is lowered, namely, the crossovers from an FO fixed point to an LM fixed point, 
and then from LM to SC. Note that, for small values of $U$, such as $U=0.025$ (black curve), 
the LM fixed point is not visible, as in this case the Kondo temperature becomes comparable 
to $\Gamma$ and $U$, and the system is close to an intermediate 
valence situation. The intriguing small dip in the impurity magnetic moment, 
as well as in the entropy (presenting a small variation with $U$), 
for temperatures in the range $10^{-2}-10^{0}$, points to the presence of van-Hove 
singularities~\cite{van-Hove}, coming from the quasi-1D band structure of the 
AGNR.

\begin{figure}[h!]
\begin{center}
\includegraphics[clip,width=3.4in]{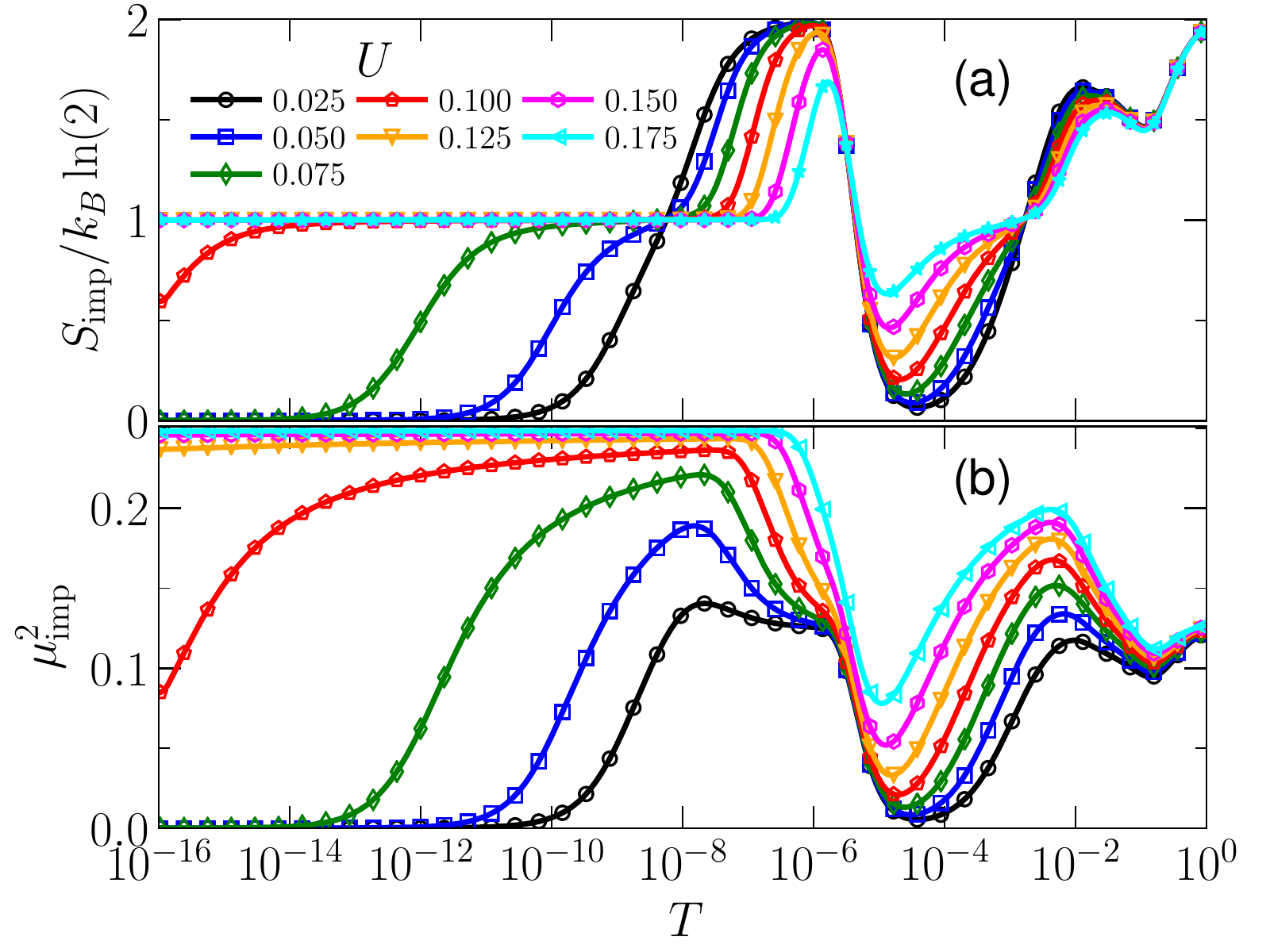}
\caption{(a) Impurity entropy $S_{\rm imp}$ and (b) magnetic moment $\mu_{\rm imp}^{2}$ for a 47-AGNR as a 
	function of temperature, for fixed RSOI induced gap $\Delta=0.9\times 10^{-5}$, $\Gamma_{\rm M}=1.0 \times 10^{-6}$, 
	and different values of $U$.} \label{fig2-AGNR}
\end{center}
\end{figure}

To see how the gap-opening introduces the reentrant SIAM behavior, discussed in Sec.~\ref{sec:model}, 
in Fig.~\ref{fig2-AGNR} we repeat the calculations shown in Fig.~\ref{fig1-AGNR}, with the same set of parameters, 
except that $\lambda_R$ is now finite, producing a gap $\Delta=0.9\times 10^{-5}$. For values 
of $U=0.025$, up to $U=0.075$, we clearly see, both from the impurity entropy $S_{\rm imp}$ [Fig.~\ref{fig2-AGNR}(a)] 
and from the impurity magnetic moment $\mu_{\rm imp}^{2}$ 
[Fig.~\ref{fig2-AGNR}(b)], the emergence of the reentrant SIAM behavior for  
temperatures below $\Delta \approx 10^{-5}$ (compare with the results in Fig.~\ref{fig1-AGNR} 
for the same temperature range). As $U$ increases, the Kondo temperature $T_{K1}$ of the first Kondo 
screening decreases, so that the unstable LM fixed point becomes more pronounced (i.e., 
extends over a larger interval of temperature). 
As a consequence, the observed decrease of $T_{K1}$, as $U$ increases, squeezes the first 
SC fixed point within a temperature 
range $\Delta \lesssim T \lesssim T_{K1}$, and, eventually, the first Kondo screening ceases to occur 
when $T_{K1}$ becomes  comparable to $\Delta$. This is manifested in the progressive enhancement 
of $S_{\rm imp}$ and $\mu_{\rm \imp}^2$ in this temperature region (because the first LM fixed point 
extends further down in temperature). It is interesting to observe that 
the reentrant Kondo temperature $T_{K2}$ decreases much more rapidly than $T_{K1}$ 
with increasing $U$, as observed in the fast increase of plateau extension of 
the reentrant LM fixed point. The decrease of $T_{K1}$ with increasing $U$ 
can be understood in terms of the Haldane expression for the Kondo temperature 
in the conventional SIAM~\cite{Haldane1978}. From our calculations we find 
that the effective Coulomb repulsion $U_{\rm eff}$ increases by increasing 
$U$ (not shown). Thus, even though the Haldane expression cannot be readily 
applied to obtain $T_{K2}$, it provide us with a good insight on why $T_{K2}$ 
decreases rapidly by increasing $U$.

\begin{figure}[t!]
\begin{center}
\includegraphics[clip,width=3.4in]{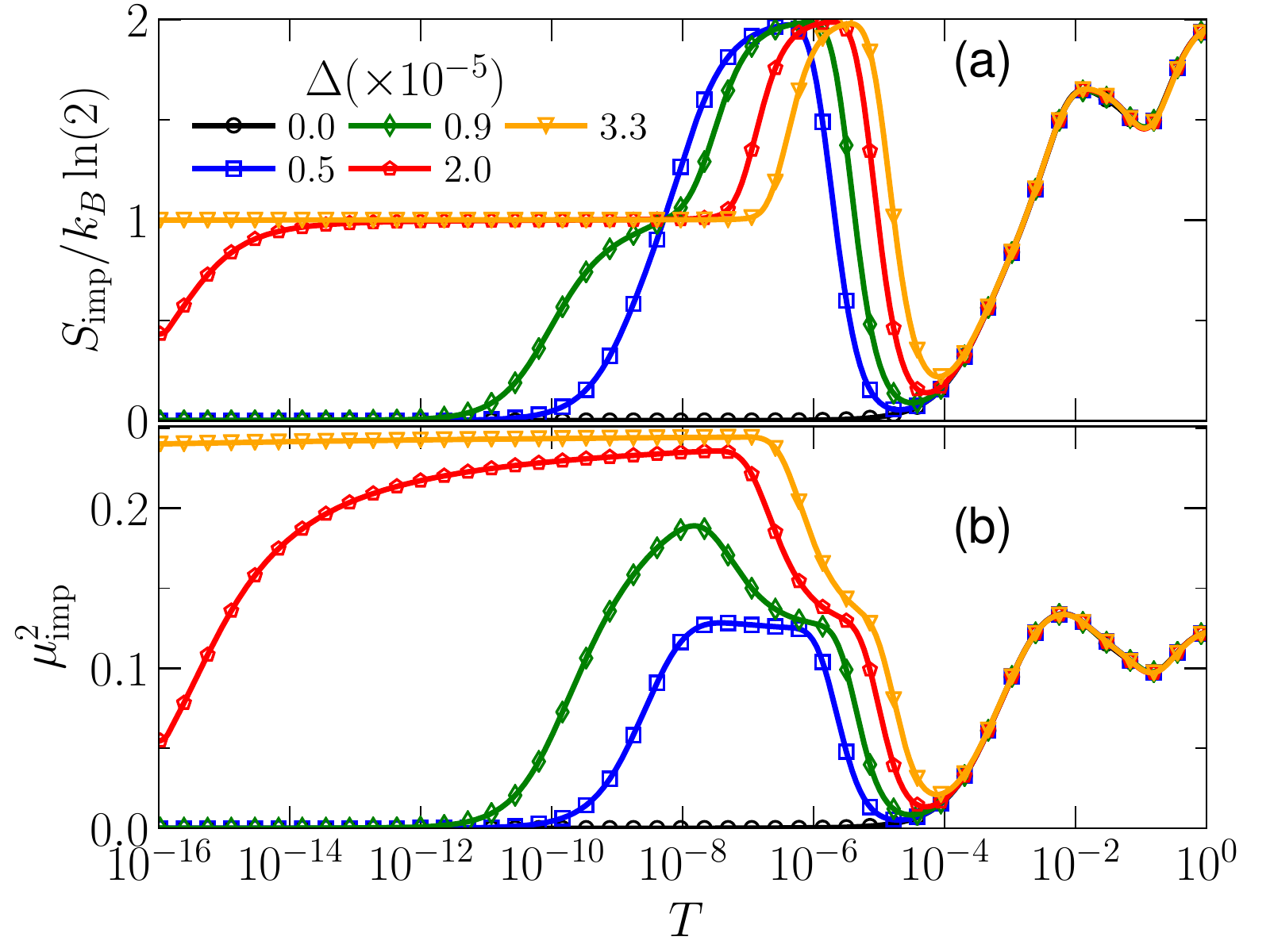}
\caption{(a) Impurity entropy $S_{\rm imp}$ and (b) magnetic moment $\mu_{\rm imp}^{2}$ for a 47-AGNR as a 
	function of temperature, for different values of RSOI induced gap 
	($0.0 \leq \Delta \leq 3.3 \times 10^{-5}$). The parameter values for both panels are 
	$\Gamma_{\rm M}=1.0\times 10^{-6}$ and $U=0.05$.}
\label{fig3-AGNR}
\end{center}
\end{figure}

Now, we proceed to a study of how the reentrant SIAM behavior is modified by changing the AGNR gap for a fixed $U$ value.   
Figs.~\ref{fig3-AGNR}(a) and \ref{fig3-AGNR}(b) show, respectively, 
$S_{\rm \imp}$ and $\mu_{\rm imp}^2$ as a function of $T$, for $U=0.05$ 
and $0 \leq \Delta \leq 3.3 \times 10^{-5}$. After interpreting the results  
in Fig.~\ref{fig2-AGNR}, as just done above, where we fixed $\Delta$ and increased $U$, the results 
in Fig.~\ref{fig3-AGNR} can be understood quite straightforwardly. Indeed, by increasing $\Delta$, the extension of the first LM fixed point 
is squeezed from below, as $T_{K1} \approx 10^{-3}$ is now fixed 
(notice the collapse of all curves, in both panels, for $T \gtrsim 10^{-4}$),  
and the extent of the first SC fixed point is determined by $\Delta$. 
In addition, the extension of the reentrant FO fixed point plateau decreases for 
increasing $\Delta$, indicating a decrease in the charge fluctuations in the reentrant SIAM for increasing $\Delta$. 
This suggests that the effective Coulomb repulsion $U_{\rm eff}$ associated 
to the reentrant SIAM increases with $\Delta$, resulting in smaller $T_{K2}$ values, which is clearly 
seen by the reentrant Kondo screening taking place at lower temperatures for larger 
$\Delta$. Moreover, for $\Delta > T_{K1}$ (not shown), no Kondo screening takes place 
as $\Delta$ exceeds $T_K$ (which is analogous to say that 
$\Gamma_{c} > \Gamma_0$) destroying the first Kondo stage, as discussed in 
Sec.~\ref{sec:model}.

An important question, mainly for experimentalists, remains to be answered, namely, 
what are the estimated values for $T_{K1}$ and $T_{K2}$ for the AGNR+QI+STM system? 
Let us first present the highest $T_{K2}$ value [blue open squares in Fig.~\ref{fig3-AGNR}(b)], 
where the Kondo temperature was obtained using Wilson's criterion, as done in Figs.~\ref{Fig_thermo_1}(b) and \ref{Fig_Thermo_gap}(b). 
We assume realistic values for the model parameters, i.e., 
nearest-neighbor hopping $t\approx 2.7$eV, which results in $D \approx 8.37$eV, thus $2\Delta=1.0\times 10^{-5}D \approx 0.08$meV, 
$U= 0.05D \approx 418$meV, and $\Gamma_{M}=1.0 \times 10^{-6}D \approx$ 8.37$\mu$eV. The NRG estimated 
values for $T_{K1}$ and $T_{K2}$ are approximately $106.72$K ($9.2$meV) and $0.5$mK ($0.043\mu$eV), respectively. 
Such a low value of $T_{K2}$ (obtained for this set of parameters) would represent an 
obstacle to the experimental detection of the reentrant Kondo physics in the AGNR+QI+STM system. 
However, notice that we have a certain degree of flexibility in varying some of the parameters, like 
the AGNR width $W$, the RSOI $\lambda_R$, (where both of them affect the $\Delta$ value), 
the coupling $\Gamma_M$ of the STM-tip to the QI, as well as its Coulomb repulsion $U$. 
In addition, based on the understanding we gathered on the physics of the reentrant Kondo, we 
have some intuition on how to increase $T_{K2}$. Indeed, the semiconducting gap $\Delta$ is located between 
$T_{K1}$ and $T_{K2}$, separated by a few orders of magnitude, i.e., $T_{K2} \ll \Delta \ll T_{K1}$, although 
there seems to be no restriction on how much $T_{K2}$ may approach $\Delta$, other 
than resulting in an unrealistically large $T_{K1}$, as both are strongly connected (see Fig.~\ref{Fig_TK2xGamma}). 
From the results in the previous sections we know that $T_{K2}$ 
should increase as $U_{\rm eff}$ decreases and $\Gamma_{\rm M}$ increases, with the former decreasing as $U$ decreases.  
Following this recipe, but still using realistic parameter values, we manage to obtain $T_{K1}=55.7$K ($4.8$meV)~\cite{hybrid} 
and $T_{K2}=10.2$mK ($0.9\mu$eV), by assuming $W\approx11.56$
nm and $\lambda_{R}=33.5$meV (resulting in $2\Delta=0.14$meV), $\Gamma_{M}=502\mu$eV, and $U=214$meV. 
This $T_{K2}$ value, we will argue below, is already much closer to being experimentally accessible.

To finish this section, without trying to exhaust the literature in the subject, 
we will place our results in the context of theoretical~\cite{Pustilnik2001,Hofstetter2004,Cornaglia2005,Zitko2010} 
and experimental~\cite{Wiel2002,Granger2005,Sasaki2009} results that are related to the occurrence of consecutive 
Kondo effects (as one lowers temperature), dubbed in the literature, in general, as two-stage Kondo effects. 
There are two distinct flavors of it: (i) in QDs containing an even number of electrons, a singlet-triplet 
Kondo effect has been observed both in vertical QDs~\cite{Sasaki2000} 
as well as in lateral QDs~\cite{Wiel2002}, and, more recently, 
in carbon nanotube QDs~\cite{Petit2014}. Consecutive Kondo effects 
(dubbed as `two-stage Kondo effect') have been observed on both sides of
the singlet-triplet transition in semiconducting QDs~\cite{Wiel2002}. 
The effects have distinct mechanisms on each side of the transition, 
and both effects require the formation of an $S=1$ state, with the 
presence of two screening channels on the triplet side and a single 
one on the singlet side. For example, in the singlet side, van der Wiel 
\emph{et al.}~\cite{Wiel2002} report values $T_{K1} \approx 3.5$K 
($300\mu$eV) and $T_{K2} \lesssim 1$K ($86\mu$eV). (ii) in double 
QD (DQD) systems, where one of the QDs (QD1) is embedded between the source and 
drain leads and the other QD (QD2) is side-coupled to QD1, through
a tunneling junction. In that case, for the right couplings between QD1 
and the Fermi sea, and between both QDs, QD1 is Kondo screened first, 
at a higher temperature $T_{K1}$, by the Fermi sea electrons. 
At a much lower temperature $T_{K2}$, QD2 will be Kondo screened 
by the quasi-particles forming the Fermi liquid ground state resulting 
from the first Kondo state. 
The spectral density that couples to QD2 is essentially the Kondo peak 
of QD1. This second flavor, although having a two-stage mechanism that is 
very diverse from the reentrant Kondo presented here, is more akin to 
our case, since $T_{K1}$ is, in general, orders of magnitude higher than 
$T_{K2}$. Therefore, the observation of its second stage has posed a 
stiff challenge to experimentalists. In that respect, it is interesting to note that 
R.~\v{Z}itko~\cite{Zitko2010}, using NRG to simulate transport properties 
of a DQD system, has claimed that Sasaki \emph{et al.}~\cite{Sasaki2009}, 
doing measurements at low temperatures (in the range of few tens of mK),
have actually observed fingerprints of the second ($T_{K2}$) Kondo stage. 
This illustrates the fact that, in our opinion, the proper use of 
gap enginnering techniques in similar systems to our AGNR+QI+STM may 
result in the observation of the second Kondo stage described here.

\section{Summary and Conclusions}\label{sec:conc}
In summary, in this paper, using Anderson's PMS and NRG approaches, we have analyzed a system involving a 
QI strongly ($\Gamma_{\rm S}$)-coupled to a semiconductor (defined by a gap $2\Delta$) 
and weakly ($\Gamma_{\rm M}$)-coupled to a metal (Fig.~\ref{fig1_new}). Our analysis 
has unveiled the existence of a sequence of two Kondo `stages': the first one, occurring at higher temperatures, 
is characterized by an unstable SC fixed point, defined by a Kondo temperature $T_{K1} > \Delta$ and associated to 
a Kondo screening that dissipates when $T \rightarrow \Delta$, 
from above. As already studied in detail in the literature (see Introduction), 
this unstable first stage Kondo may not happen at all in case $\Gamma_0=\Gamma_{\rm S} + \Gamma_{\rm M} < \Gamma_c$, as 
discussed at the beginning of Sec.~\ref{sec:model} (see Fig.~\ref{PMS}). In case it does happen, it will be 
followed, for $T \lesssim \Delta$, by a second stage Kondo, characterized by a Kondo temperature $T_{K2} \ll T_{K1}$, 
that presents a \emph{replica} of the usual SIAM-fixed-points sequence (FO $\rightarrow$ LM $\rightarrow$ SC), but 
for which, in contrast to the first stage Kondo, the SC fixed point is now stable. We dub this `emergent' SIAM as 
\emph{reentrant} effective SIAM, with an effective Hubbard $U_{\rm eff} \ll U$, which is clearly displayed as a peak in 
the impurity LDOS, alongside a second Kondo peak (see Fig.~\ref{Fig_DOS}). The properties of both stages are 
thoroughly analyzed through the impurity's thermodynamic properties and LDOS, using NRG. The intuitive picture that 
emerges, after the analysis of the NRG results, is a simple one: the high temperature first Kondo state develops through 
impurity screening by thermally excited semiconducting electrons, while the second stage involves 
screening by metallic electrons, once the semiconducting 
electrons are out of reach to thermal excitations ($T < \Delta$) and only the metallic (low) spectral weight 
inside the gap is available for impurity screening. In addition, 
in Sec.~\ref{sec:AGNR}, we propose a realistic system where the reentrant Kondo stage may possibly be experimentally observed: 
a magnetic impurity strongly coupled to an AGNR and weakly coupled to an STM tip. 
The proposal is based on the use of an electric-field-induced 
RSOI to tune a gap $2\Delta$ in an otherwise metallic AGNR, and, through a full NRG analysis of this system, using 
realistic parameters, we show that both stages may be considered as experimentally accessible, as a 
recent theory work~\cite{Zitko2010} has suggested that the second stage Kondo, expected in DQD systems, has actually been 
observed~\cite{Sasaki2009} through charge transport measurements at low temperatures in a semiconducting DQD system. 
We hope that our findings may spur theory groups to apply other techniques to the analysis of this model, as well 
as study its charge transport properties, which is the preferred experimental tool for spectroscopic analysis of these 
mesoscopic systems. We also expect to motivate the proposal of additional systems that could be similarly modeled, involving 
not only carbon materials (as we have proposed), but also containing related materials that are amenable to appropriate 
gap engineering. 

\section{Acknowledgments}

It is a pleasure to acknowledge fruitful discussions with G.J.~Ferreira. 
GD acknowledges an MS scholarship from the Brazilian agency Coordena\c{c}\~ao 
de Aperfei\c{c}oamento de Pessoal de N\'ivel Superior (CAPES), EV acknowledges support from Ohio 
University within the Robert Glidden Visiting Professorship program. Additional 
support from the Brazilian funding agencies CAPES and FAPEMIG is also acknowledged. 

\bibliography{references}

%merlin.mbs apsrev4-1.bst 2010-07-25 4.21a (PWD, AO, DPC) hacked
%Control: key (0)
%Control: author (72) initials jnrlst
%Control: editor formatted (1) identically to author
%Control: production of article title (-1) disabled
%Control: page (0) single
%Control: year (1) truncated
%Control: production of eprint (0) enabled
\begin{thebibliography}{81}%
\makeatletter
\providecommand \@ifxundefined [1]{%
 \@ifx{#1\undefined}
}%
\providecommand \@ifnum [1]{%
 \ifnum #1\expandafter \@firstoftwo
 \else \expandafter \@secondoftwo
 \fi
}%
\providecommand \@ifx [1]{%
 \ifx #1\expandafter \@firstoftwo
 \else \expandafter \@secondoftwo
 \fi
}%
\providecommand \natexlab [1]{#1}%
\providecommand \enquote  [1]{``#1''}%
\providecommand \bibnamefont  [1]{#1}%
\providecommand \bibfnamefont [1]{#1}%
\providecommand \citenamefont [1]{#1}%
\providecommand \href@noop [0]{\@secondoftwo}%
\providecommand \href [0]{\begingroup \@sanitize@url \@href}%
\providecommand \@href[1]{\@@startlink{#1}\@@href}%
\providecommand \@@href[1]{\endgroup#1\@@endlink}%
\providecommand \@sanitize@url [0]{\catcode `\\12\catcode `\$12\catcode
  `\&12\catcode `\#12\catcode `\^12\catcode `\_12\catcode `\%12\relax}%
\providecommand \@@startlink[1]{}%
\providecommand \@@endlink[0]{}%
\providecommand \url  [0]{\begingroup\@sanitize@url \@url }%
\providecommand \@url [1]{\endgroup\@href {#1}{\urlprefix }}%
\providecommand \urlprefix  [0]{URL }%
\providecommand \Eprint [0]{\href }%
\providecommand \doibase [0]{http://dx.doi.org/}%
\providecommand \selectlanguage [0]{\@gobble}%
\providecommand \bibinfo  [0]{\@secondoftwo}%
\providecommand \bibfield  [0]{\@secondoftwo}%
\providecommand \translation [1]{[#1]}%
\providecommand \BibitemOpen [0]{}%
\providecommand \bibitemStop [0]{}%
\providecommand \bibitemNoStop [0]{.\EOS\space}%
\providecommand \EOS [0]{\spacefactor3000\relax}%
\providecommand \BibitemShut  [1]{\csname bibitem#1\endcsname}%
\let\auto@bib@innerbib\@empty
%</preamble>
\bibitem [{\citenamefont {Anderson}(1972)}]{Anderson393}%
  \BibitemOpen
  \bibfield  {author} {\bibinfo {author} {\bibfnamefont {P.~W.}\ \bibnamefont
  {Anderson}},\ }\href {\doibase 10.1126/science.177.4047.393} {\bibfield
  {journal} {\bibinfo  {journal} {Science}\ }\textbf {\bibinfo {volume}
  {177}},\ \bibinfo {pages} {393} (\bibinfo {year} {1972})}\BibitemShut
  {NoStop}%
\bibitem [{\citenamefont {Kondo}(1964)}]{Kondo1964}%
  \BibitemOpen
  \bibfield  {author} {\bibinfo {author} {\bibfnamefont {J.}~\bibnamefont
  {Kondo}},\ }\href {\doibase 10.1143/PTP.32.37} {\bibfield  {journal}
  {\bibinfo  {journal} {Prog. Theor. Phys.}\ }\textbf {\bibinfo {volume}
  {32}},\ \bibinfo {pages} {37} (\bibinfo {year} {1964})}\BibitemShut {NoStop}%
\bibitem [{\citenamefont {Hewson}(1993)}]{Hewson-Kondo}%
  \BibitemOpen
  \bibfield  {author} {\bibinfo {author} {\bibfnamefont {A.~C.}\ \bibnamefont
  {Hewson}},\ }\href@noop {} {\emph {\bibinfo {title} {The Kondo problem to
  heavy fermions}}}\ (\bibinfo  {publisher} {Cambridge University Press},\
  \bibinfo {year} {1993})\BibitemShut {NoStop}%
\bibitem [{\citenamefont {Goldhaber-Gordon}\ \emph {et~al.}(1998)\citenamefont
  {Goldhaber-Gordon}, \citenamefont {Shtrikman}, \citenamefont {Mahalu},
  \citenamefont {Abusch-Magder}, \citenamefont {Meirav},\ and\ \citenamefont
  {Kastner}}]{Goldhaber-Gordon1998}%
  \BibitemOpen
  \bibfield  {author} {\bibinfo {author} {\bibfnamefont {D.}~\bibnamefont
  {Goldhaber-Gordon}}, \bibinfo {author} {\bibfnamefont {H.}~\bibnamefont
  {Shtrikman}}, \bibinfo {author} {\bibfnamefont {D.}~\bibnamefont {Mahalu}},
  \bibinfo {author} {\bibfnamefont {D.}~\bibnamefont {Abusch-Magder}}, \bibinfo
  {author} {\bibfnamefont {U.}~\bibnamefont {Meirav}}, \ and\ \bibinfo {author}
  {\bibfnamefont {M.}~\bibnamefont {Kastner}},\ }\href {\doibase
  10.1103/10.1038/34373} {\bibfield  {journal} {\bibinfo  {journal} {Nature}\
  }\textbf {\bibinfo {volume} {391}},\ \bibinfo {pages} {156} (\bibinfo {year}
  {1998})}\BibitemShut {NoStop}%
\bibitem [{\citenamefont {Anderson}(1961)}]{Anderson1961}%
  \BibitemOpen
  \bibfield  {author} {\bibinfo {author} {\bibfnamefont {P.~W.}\ \bibnamefont
  {Anderson}},\ }\href {\doibase 10.1103/PhysRev.124.41} {\bibfield  {journal}
  {\bibinfo  {journal} {Phys. Rev.}\ }\textbf {\bibinfo {volume} {124}},\
  \bibinfo {pages} {41} (\bibinfo {year} {1961})}\BibitemShut {NoStop}%
\bibitem [{\citenamefont {Krishna-murthy}\ \emph {et~al.}(1980)\citenamefont
  {Krishna-murthy}, \citenamefont {Wilkins},\ and\ \citenamefont
  {Wilson}}]{Krishna-murthy1980}%
  \BibitemOpen
  \bibfield  {author} {\bibinfo {author} {\bibfnamefont {H.~R.}\ \bibnamefont
  {Krishna-murthy}}, \bibinfo {author} {\bibfnamefont {J.~W.}\ \bibnamefont
  {Wilkins}}, \ and\ \bibinfo {author} {\bibfnamefont {K.~G.}\ \bibnamefont
  {Wilson}},\ }\href {\doibase 10.1103/PhysRevB.21.1003} {\bibfield  {journal}
  {\bibinfo  {journal} {Phys. Rev. B}\ }\textbf {\bibinfo {volume} {21}},\
  \bibinfo {pages} {1003} (\bibinfo {year} {1980})}\BibitemShut {NoStop}%
\bibitem [{\citenamefont {Bulla}\ \emph {et~al.}(2008)\citenamefont {Bulla},
  \citenamefont {Costi},\ and\ \citenamefont {Pruschke}}]{Bulla2008}%
  \BibitemOpen
  \bibfield  {author} {\bibinfo {author} {\bibfnamefont {R.}~\bibnamefont
  {Bulla}}, \bibinfo {author} {\bibfnamefont {T.~A.}\ \bibnamefont {Costi}}, \
  and\ \bibinfo {author} {\bibfnamefont {T.}~\bibnamefont {Pruschke}},\ }\href
  {\doibase 10.1103/RevModPhys.80.395} {\bibfield  {journal} {\bibinfo
  {journal} {Rev. Mod. Phys.}\ }\textbf {\bibinfo {volume} {80}},\ \bibinfo
  {pages} {395} (\bibinfo {year} {2008})}\BibitemShut {NoStop}%
\bibitem [{\citenamefont {Takegahara}\ \emph {et~al.}(1992)\citenamefont
  {Takegahara}, \citenamefont {Shimizu},\ and\ \citenamefont
  {Sakai}}]{Takegahara1992}%
  \BibitemOpen
  \bibfield  {author} {\bibinfo {author} {\bibfnamefont {K.}~\bibnamefont
  {Takegahara}}, \bibinfo {author} {\bibfnamefont {Y.}~\bibnamefont {Shimizu}},
  \ and\ \bibinfo {author} {\bibfnamefont {O.}~\bibnamefont {Sakai}},\ }\href
  {\doibase 10.1143/JPSJ.61.3443} {\bibfield  {journal} {\bibinfo  {journal}
  {J. Phys. Soc. Jpn.}\ }\textbf {\bibinfo {volume} {61}},\ \bibinfo {pages}
  {3443} (\bibinfo {year} {1992})}\BibitemShut {NoStop}%
\bibitem [{\citenamefont {Takegahara}\ \emph {et~al.}(1993)\citenamefont
  {Takegahara}, \citenamefont {Shimizu}, \citenamefont {Goto},\ and\
  \citenamefont {Sakai}}]{Takegahara1993}%
  \BibitemOpen
  \bibfield  {author} {\bibinfo {author} {\bibfnamefont {K.}~\bibnamefont
  {Takegahara}}, \bibinfo {author} {\bibfnamefont {Y.}~\bibnamefont {Shimizu}},
  \bibinfo {author} {\bibfnamefont {N.}~\bibnamefont {Goto}}, \ and\ \bibinfo
  {author} {\bibfnamefont {O.}~\bibnamefont {Sakai}},\ }\href {\doibase
  10.1016/0921-4526(93)90579-U} {\bibfield  {journal} {\bibinfo  {journal}
  {Physica B}\ }\textbf {\bibinfo {volume} {186}},\ \bibinfo {pages} {381}
  (\bibinfo {year} {1993})}\BibitemShut {NoStop}%
\bibitem [{\citenamefont {Saso}(1992)}]{Saso1992}%
  \BibitemOpen
  \bibfield  {author} {\bibinfo {author} {\bibfnamefont {T.}~\bibnamefont
  {Saso}},\ }\href {\doibase 10.1143/JPSJ.61.3439} {\bibfield  {journal}
  {\bibinfo  {journal} {J. Phys. Soc. Jpn.}\ }\textbf {\bibinfo {volume}
  {61}},\ \bibinfo {pages} {3439} (\bibinfo {year} {1992})}\BibitemShut
  {NoStop}%
\bibitem [{\citenamefont {Ogura}\ and\ \citenamefont {Saso}(1993)}]{Ogura1993}%
  \BibitemOpen
  \bibfield  {author} {\bibinfo {author} {\bibfnamefont {J.}~\bibnamefont
  {Ogura}}\ and\ \bibinfo {author} {\bibfnamefont {T.}~\bibnamefont {Saso}},\
  }\href {\doibase 10.1143/JPSJ.62.4364} {\bibfield  {journal} {\bibinfo
  {journal} {J. Phys. Soc. Jpn.}\ }\textbf {\bibinfo {volume} {62}},\ \bibinfo
  {pages} {4364} (\bibinfo {year} {1993})}\BibitemShut {NoStop}%
\bibitem [{\citenamefont {Cruz}\ \emph {et~al.}(1995)\citenamefont {Cruz},
  \citenamefont {Phillips},\ and\ \citenamefont {Castro~Neto}}]{Cruz1995}%
  \BibitemOpen
  \bibfield  {author} {\bibinfo {author} {\bibfnamefont {L.}~\bibnamefont
  {Cruz}}, \bibinfo {author} {\bibfnamefont {P.}~\bibnamefont {Phillips}}, \
  and\ \bibinfo {author} {\bibfnamefont {A.~H.}\ \bibnamefont {Castro~Neto}},\
  }\href {\doibase 10.1209/0295-5075/29/5/007} {\bibfield  {journal} {\bibinfo
  {journal} {EPL}\ }\textbf {\bibinfo {volume} {29}},\ \bibinfo {pages} {389}
  (\bibinfo {year} {1995})}\BibitemShut {NoStop}%
\bibitem [{\citenamefont {Yu}\ and\ \citenamefont {Guerrero}(1996)}]{Yu1996}%
  \BibitemOpen
  \bibfield  {author} {\bibinfo {author} {\bibfnamefont {C.~C.}\ \bibnamefont
  {Yu}}\ and\ \bibinfo {author} {\bibfnamefont {M.}~\bibnamefont {Guerrero}},\
  }\href {\doibase 10.1103/PhysRevB.54.8556} {\bibfield  {journal} {\bibinfo
  {journal} {Phys. Rev. B}\ }\textbf {\bibinfo {volume} {54}},\ \bibinfo
  {pages} {8556} (\bibinfo {year} {1996})}\BibitemShut {NoStop}%
\bibitem [{\citenamefont {Chen}\ and\ \citenamefont
  {Jayaprakash}(1998)}]{Chen1998}%
  \BibitemOpen
  \bibfield  {author} {\bibinfo {author} {\bibfnamefont {K.}~\bibnamefont
  {Chen}}\ and\ \bibinfo {author} {\bibfnamefont {C.}~\bibnamefont
  {Jayaprakash}},\ }\href {\doibase 10.1103/PhysRevB.57.5225} {\bibfield
  {journal} {\bibinfo  {journal} {Phys. Rev. B}\ }\textbf {\bibinfo {volume}
  {57}},\ \bibinfo {pages} {5225} (\bibinfo {year} {1998})}\BibitemShut
  {NoStop}%
\bibitem [{\citenamefont {Moca}\ and\ \citenamefont {Roman}(2010)}]{Moca2010}%
  \BibitemOpen
  \bibfield  {author} {\bibinfo {author} {\bibfnamefont {C.~P.}\ \bibnamefont
  {Moca}}\ and\ \bibinfo {author} {\bibfnamefont {A.}~\bibnamefont {Roman}},\
  }\href {\doibase 10.1103/PhysRevB.81.235106} {\bibfield  {journal} {\bibinfo
  {journal} {Phys. Rev. B}\ }\textbf {\bibinfo {volume} {81}},\ \bibinfo
  {pages} {235106} (\bibinfo {year} {2010})}\BibitemShut {NoStop}%
\bibitem [{\citenamefont {Galpin}\ and\ \citenamefont
  {Logan}(2008{\natexlab{a}})}]{Galpin2008a}%
  \BibitemOpen
  \bibfield  {author} {\bibinfo {author} {\bibfnamefont {M.~R.}\ \bibnamefont
  {Galpin}}\ and\ \bibinfo {author} {\bibfnamefont {D.~E.}\ \bibnamefont
  {Logan}},\ }\href {\doibase 10.1103/PhysRevB.77.195108} {\bibfield  {journal}
  {\bibinfo  {journal} {Phys. Rev. B}\ }\textbf {\bibinfo {volume} {77}},\
  \bibinfo {pages} {195108} (\bibinfo {year} {2008}{\natexlab{a}})}\BibitemShut
  {NoStop}%
\bibitem [{\citenamefont {Galpin}\ and\ \citenamefont
  {Logan}(2008{\natexlab{b}})}]{Galpin2008b}%
  \BibitemOpen
  \bibfield  {author} {\bibinfo {author} {\bibfnamefont {M.~R.}\ \bibnamefont
  {Galpin}}\ and\ \bibinfo {author} {\bibfnamefont {D.~E.}\ \bibnamefont
  {Logan}},\ }\href {\doibase 10.1140/epjb/e2008-00138-5} {\bibfield  {journal}
  {\bibinfo  {journal} {Eur. Phys. J. B}\ }\textbf {\bibinfo {volume} {62}},\
  \bibinfo {pages} {129} (\bibinfo {year} {2008}{\natexlab{b}})}\BibitemShut
  {NoStop}%
\bibitem [{bil()}]{bilayer}%
  \BibitemOpen
  \href@noop {} {}\bibinfo {note} {An alternate proposal could explore the
  \emph{Mexican hat} electronic dispersion obtained in biased bilayer graphene,
  where the gap can be modified by an external electric field, as
  experimentally reported in Castro \emph{et al}., Phys. Rev. Lett. 99, 216802
  (2007).}\BibitemShut {Stop}%
\bibitem [{\citenamefont {Nygard}\ \emph {et~al.}(2000)\citenamefont {Nygard},
  \citenamefont {Cobden},\ and\ \citenamefont {Lindelof}}]{Nygard}%
  \BibitemOpen
  \bibfield  {author} {\bibinfo {author} {\bibfnamefont {J.}~\bibnamefont
  {Nygard}}, \bibinfo {author} {\bibfnamefont {D.~H.}\ \bibnamefont {Cobden}},
  \ and\ \bibinfo {author} {\bibfnamefont {P.~E.}\ \bibnamefont {Lindelof}},\
  }\href {\doibase 10.1038/35042545} {\bibfield  {journal} {\bibinfo  {journal}
  {Nature}\ }\textbf {\bibinfo {volume} {408}},\ \bibinfo {pages} {342}
  (\bibinfo {year} {2000})}\BibitemShut {NoStop}%
\bibitem [{\citenamefont {Jarillo-Herrero}\ \emph {et~al.}(2005)\citenamefont
  {Jarillo-Herrero}, \citenamefont {Kong}, \citenamefont {van~der Zant},
  \citenamefont {Dekker}, \citenamefont {Kouwenhoven},\ and\ \citenamefont
  {De~Franceschi}}]{Jarillo}%
  \BibitemOpen
  \bibfield  {author} {\bibinfo {author} {\bibfnamefont {P.}~\bibnamefont
  {Jarillo-Herrero}}, \bibinfo {author} {\bibfnamefont {J.}~\bibnamefont
  {Kong}}, \bibinfo {author} {\bibfnamefont {H.~S.}\ \bibnamefont {van~der
  Zant}}, \bibinfo {author} {\bibfnamefont {C.}~\bibnamefont {Dekker}},
  \bibinfo {author} {\bibfnamefont {L.~P.}\ \bibnamefont {Kouwenhoven}}, \ and\
  \bibinfo {author} {\bibfnamefont {S.}~\bibnamefont {De~Franceschi}},\ }\href
  {\doibase 10.1038/nature03422} {\bibfield  {journal} {\bibinfo  {journal}
  {Nature}\ }\textbf {\bibinfo {volume} {434}},\ \bibinfo {pages} {484}
  (\bibinfo {year} {2005})}\BibitemShut {NoStop}%
\bibitem [{\citenamefont {Sengupta}\ and\ \citenamefont
  {Baskaran}(2008)}]{PhysRevB.77.045417}%
  \BibitemOpen
  \bibfield  {author} {\bibinfo {author} {\bibfnamefont {K.}~\bibnamefont
  {Sengupta}}\ and\ \bibinfo {author} {\bibfnamefont {G.}~\bibnamefont
  {Baskaran}},\ }\href {\doibase 10.1103/PhysRevB.77.045417} {\bibfield
  {journal} {\bibinfo  {journal} {Phys. Rev. B}\ }\textbf {\bibinfo {volume}
  {77}},\ \bibinfo {pages} {045417} (\bibinfo {year} {2008})}\BibitemShut
  {NoStop}%
\bibitem [{\citenamefont {Chao}\ and\ \citenamefont
  {Aji}(2011)}]{PhysRevB.83.165449}%
  \BibitemOpen
  \bibfield  {author} {\bibinfo {author} {\bibfnamefont {S.-P.}\ \bibnamefont
  {Chao}}\ and\ \bibinfo {author} {\bibfnamefont {V.}~\bibnamefont {Aji}},\
  }\href {\doibase 10.1103/PhysRevB.83.165449} {\bibfield  {journal} {\bibinfo
  {journal} {Phys. Rev. B}\ }\textbf {\bibinfo {volume} {83}},\ \bibinfo
  {pages} {165449} (\bibinfo {year} {2011})}\BibitemShut {NoStop}%
\bibitem [{\citenamefont {Fritz}\ and\ \citenamefont
  {Vojta}(2013)}]{0034-4885-76-3-032501}%
  \BibitemOpen
  \bibfield  {author} {\bibinfo {author} {\bibfnamefont {L.}~\bibnamefont
  {Fritz}}\ and\ \bibinfo {author} {\bibfnamefont {M.}~\bibnamefont {Vojta}},\
  }\href {http://stacks.iop.org/0034-4885/76/i=3/a=032501} {\bibfield
  {journal} {\bibinfo  {journal} {Rep. Prog. Phys.}\ }\textbf {\bibinfo
  {volume} {76}},\ \bibinfo {pages} {032501} (\bibinfo {year}
  {2013})}\BibitemShut {NoStop}%
\bibitem [{\citenamefont {Zhu}\ and\ \citenamefont
  {Berakdar}(2011)}]{PhysRevB.84.165105}%
  \BibitemOpen
  \bibfield  {author} {\bibinfo {author} {\bibfnamefont {Z.-G.}\ \bibnamefont
  {Zhu}}\ and\ \bibinfo {author} {\bibfnamefont {J.}~\bibnamefont {Berakdar}},\
  }\href {\doibase 10.1103/PhysRevB.84.165105} {\bibfield  {journal} {\bibinfo
  {journal} {Phys. Rev. B}\ }\textbf {\bibinfo {volume} {84}},\ \bibinfo
  {pages} {165105} (\bibinfo {year} {2011})}\BibitemShut {NoStop}%
\bibitem [{\citenamefont {Kharitonov}\ and\ \citenamefont
  {Kotliar}(2013)}]{PhysRevB.88.201103}%
  \BibitemOpen
  \bibfield  {author} {\bibinfo {author} {\bibfnamefont {M.}~\bibnamefont
  {Kharitonov}}\ and\ \bibinfo {author} {\bibfnamefont {G.}~\bibnamefont
  {Kotliar}},\ }\href {\doibase 10.1103/PhysRevB.88.201103} {\bibfield
  {journal} {\bibinfo  {journal} {Phys. Rev. B}\ }\textbf {\bibinfo {volume}
  {88}},\ \bibinfo {pages} {201103} (\bibinfo {year} {2013})}\BibitemShut
  {NoStop}%
\bibitem [{\citenamefont {Fang}\ and\ \citenamefont
  {Sun}(2013)}]{PhysRevB.87.075116}%
  \BibitemOpen
  \bibfield  {author} {\bibinfo {author} {\bibfnamefont {T.-F.}\ \bibnamefont
  {Fang}}\ and\ \bibinfo {author} {\bibfnamefont {Q.-f.}\ \bibnamefont {Sun}},\
  }\href {\doibase 10.1103/PhysRevB.87.075116} {\bibfield  {journal} {\bibinfo
  {journal} {Phys. Rev. B}\ }\textbf {\bibinfo {volume} {87}},\ \bibinfo
  {pages} {075116} (\bibinfo {year} {2013})}\BibitemShut {NoStop}%
\bibitem [{\citenamefont {Mastrogiuseppe}\ \emph {et~al.}(2014)\citenamefont
  {Mastrogiuseppe}, \citenamefont {Wong}, \citenamefont {Ingersent},
  \citenamefont {Ulloa},\ and\ \citenamefont {Sandler}}]{PhysRevB.90.035426}%
  \BibitemOpen
  \bibfield  {author} {\bibinfo {author} {\bibfnamefont {D.}~\bibnamefont
  {Mastrogiuseppe}}, \bibinfo {author} {\bibfnamefont {A.}~\bibnamefont
  {Wong}}, \bibinfo {author} {\bibfnamefont {K.}~\bibnamefont {Ingersent}},
  \bibinfo {author} {\bibfnamefont {S.~E.}\ \bibnamefont {Ulloa}}, \ and\
  \bibinfo {author} {\bibfnamefont {N.}~\bibnamefont {Sandler}},\ }\href
  {\doibase 10.1103/PhysRevB.90.035426} {\bibfield  {journal} {\bibinfo
  {journal} {Phys. Rev. B}\ }\textbf {\bibinfo {volume} {90}},\ \bibinfo
  {pages} {035426} (\bibinfo {year} {2014})}\BibitemShut {NoStop}%
\bibitem [{\citenamefont {Li}\ \emph {et~al.}(2019)\citenamefont {Li},
  \citenamefont {Fang}, \citenamefont {Guo},\ and\ \citenamefont
  {Sun}}]{PhysRevB.100.115115}%
  \BibitemOpen
  \bibfield  {author} {\bibinfo {author} {\bibfnamefont {G.-Y.}\ \bibnamefont
  {Li}}, \bibinfo {author} {\bibfnamefont {T.-F.}\ \bibnamefont {Fang}},
  \bibinfo {author} {\bibfnamefont {A.-M.}\ \bibnamefont {Guo}}, \ and\
  \bibinfo {author} {\bibfnamefont {Q.-F.}\ \bibnamefont {Sun}},\ }\href
  {\doibase 10.1103/PhysRevB.100.115115} {\bibfield  {journal} {\bibinfo
  {journal} {Phys. Rev. B}\ }\textbf {\bibinfo {volume} {100}},\ \bibinfo
  {pages} {115115} (\bibinfo {year} {2019})}\BibitemShut {NoStop}%
\bibitem [{\citenamefont {Haase}\ \emph {et~al.}(2011)\citenamefont {Haase},
  \citenamefont {Fuchs}, \citenamefont {Pruschke}, \citenamefont {Ochoa},\ and\
  \citenamefont {Guinea}}]{PhysRevB.83.241408}%
  \BibitemOpen
  \bibfield  {author} {\bibinfo {author} {\bibfnamefont {P.}~\bibnamefont
  {Haase}}, \bibinfo {author} {\bibfnamefont {S.}~\bibnamefont {Fuchs}},
  \bibinfo {author} {\bibfnamefont {T.}~\bibnamefont {Pruschke}}, \bibinfo
  {author} {\bibfnamefont {H.}~\bibnamefont {Ochoa}}, \ and\ \bibinfo {author}
  {\bibfnamefont {F.}~\bibnamefont {Guinea}},\ }\href {\doibase
  10.1103/PhysRevB.83.241408} {\bibfield  {journal} {\bibinfo  {journal} {Phys.
  Rev. B}\ }\textbf {\bibinfo {volume} {83}},\ \bibinfo {pages} {241408}
  (\bibinfo {year} {2011})}\BibitemShut {NoStop}%
\bibitem [{\citenamefont {Mitchell}\ and\ \citenamefont
  {Fritz}(2013)}]{PhysRevB.88.075104}%
  \BibitemOpen
  \bibfield  {author} {\bibinfo {author} {\bibfnamefont {A.~K.}\ \bibnamefont
  {Mitchell}}\ and\ \bibinfo {author} {\bibfnamefont {L.}~\bibnamefont
  {Fritz}},\ }\href {\doibase 10.1103/PhysRevB.88.075104} {\bibfield  {journal}
  {\bibinfo  {journal} {Phys. Rev. B}\ }\textbf {\bibinfo {volume} {88}},\
  \bibinfo {pages} {075104} (\bibinfo {year} {2013})}\BibitemShut {NoStop}%
\bibitem [{\citenamefont {May}\ \emph {et~al.}(2018)\citenamefont {May},
  \citenamefont {Lo}, \citenamefont {Deltenre}, \citenamefont {Henke},
  \citenamefont {Mao}, \citenamefont {Jiang}, \citenamefont {Li}, \citenamefont
  {Andrei}, \citenamefont {Guo},\ and\ \citenamefont
  {Anders}}]{PhysRevB.97.155419}%
  \BibitemOpen
  \bibfield  {author} {\bibinfo {author} {\bibfnamefont {D.}~\bibnamefont
  {May}}, \bibinfo {author} {\bibfnamefont {P.-W.}\ \bibnamefont {Lo}},
  \bibinfo {author} {\bibfnamefont {K.}~\bibnamefont {Deltenre}}, \bibinfo
  {author} {\bibfnamefont {A.}~\bibnamefont {Henke}}, \bibinfo {author}
  {\bibfnamefont {J.}~\bibnamefont {Mao}}, \bibinfo {author} {\bibfnamefont
  {Y.}~\bibnamefont {Jiang}}, \bibinfo {author} {\bibfnamefont
  {G.}~\bibnamefont {Li}}, \bibinfo {author} {\bibfnamefont {E.~Y.}\
  \bibnamefont {Andrei}}, \bibinfo {author} {\bibfnamefont {G.-Y.}\
  \bibnamefont {Guo}}, \ and\ \bibinfo {author} {\bibfnamefont {F.~B.}\
  \bibnamefont {Anders}},\ }\href {\doibase 10.1103/PhysRevB.97.155419}
  {\bibfield  {journal} {\bibinfo  {journal} {Phys. Rev. B}\ }\textbf {\bibinfo
  {volume} {97}},\ \bibinfo {pages} {155419} (\bibinfo {year}
  {2018})}\BibitemShut {NoStop}%
\bibitem [{\citenamefont {Jiang}\ \emph {et~al.}(2018)\citenamefont {Jiang},
  \citenamefont {Lo}, \citenamefont {May}, \citenamefont {Li}, \citenamefont
  {Guo}, \citenamefont {Anders}, \citenamefont {Taniguchi}, \citenamefont
  {Watanabe}, \citenamefont {Mao},\ and\ \citenamefont {Andrei}}]{Jiang2018}%
  \BibitemOpen
  \bibfield  {author} {\bibinfo {author} {\bibfnamefont {Y.}~\bibnamefont
  {Jiang}}, \bibinfo {author} {\bibfnamefont {P.-W.}\ \bibnamefont {Lo}},
  \bibinfo {author} {\bibfnamefont {D.}~\bibnamefont {May}}, \bibinfo {author}
  {\bibfnamefont {G.}~\bibnamefont {Li}}, \bibinfo {author} {\bibfnamefont
  {G.-Y.}\ \bibnamefont {Guo}}, \bibinfo {author} {\bibfnamefont {F.~B.}\
  \bibnamefont {Anders}}, \bibinfo {author} {\bibfnamefont {T.}~\bibnamefont
  {Taniguchi}}, \bibinfo {author} {\bibfnamefont {K.}~\bibnamefont {Watanabe}},
  \bibinfo {author} {\bibfnamefont {J.}~\bibnamefont {Mao}}, \ and\ \bibinfo
  {author} {\bibfnamefont {E.~Y.}\ \bibnamefont {Andrei}},\ }\href {\doibase
  10.1038/s41467-018-04812-6} {\bibfield  {journal} {\bibinfo  {journal} {Nat.
  Commun.}\ }\textbf {\bibinfo {volume} {9}},\ \bibinfo {pages} {2349}
  (\bibinfo {year} {2018})}\BibitemShut {NoStop}%
\bibitem [{\citenamefont {Li}\ \emph {et~al.}(2013)\citenamefont {Li},
  \citenamefont {Ni}, \citenamefont {Zhong}, \citenamefont {Fang},\ and\
  \citenamefont {Luo}}]{Li_2013}%
  \BibitemOpen
  \bibfield  {author} {\bibinfo {author} {\bibfnamefont {L.}~\bibnamefont
  {Li}}, \bibinfo {author} {\bibfnamefont {Y.-Y.}\ \bibnamefont {Ni}}, \bibinfo
  {author} {\bibfnamefont {Y.}~\bibnamefont {Zhong}}, \bibinfo {author}
  {\bibfnamefont {T.-F.}\ \bibnamefont {Fang}}, \ and\ \bibinfo {author}
  {\bibfnamefont {H.-G.}\ \bibnamefont {Luo}},\ }\href {\doibase
  10.1088/1367-2630/15/5/053018} {\bibfield  {journal} {\bibinfo  {journal}
  {New J. Phys.}\ }\textbf {\bibinfo {volume} {15}},\ \bibinfo {pages} {053018}
  (\bibinfo {year} {2013})}\BibitemShut {NoStop}%
\bibitem [{\citenamefont {Ren}\ \emph {et~al.}(2014)\citenamefont {Ren},
  \citenamefont {Guo}, \citenamefont {Pan}, \citenamefont {Zhang},
  \citenamefont {Wu}, \citenamefont {Luo}, \citenamefont {Du}, \citenamefont
  {Pantelides},\ and\ \citenamefont {Gao}}]{nl501425n}%
  \BibitemOpen
  \bibfield  {author} {\bibinfo {author} {\bibfnamefont {J.}~\bibnamefont
  {Ren}}, \bibinfo {author} {\bibfnamefont {H.}~\bibnamefont {Guo}}, \bibinfo
  {author} {\bibfnamefont {J.}~\bibnamefont {Pan}}, \bibinfo {author}
  {\bibfnamefont {Y.~Y.}\ \bibnamefont {Zhang}}, \bibinfo {author}
  {\bibfnamefont {X.}~\bibnamefont {Wu}}, \bibinfo {author} {\bibfnamefont
  {H.-G.}\ \bibnamefont {Luo}}, \bibinfo {author} {\bibfnamefont
  {S.}~\bibnamefont {Du}}, \bibinfo {author} {\bibfnamefont {S.~T.}\
  \bibnamefont {Pantelides}}, \ and\ \bibinfo {author} {\bibfnamefont {H.-J.}\
  \bibnamefont {Gao}},\ }\href {\doibase 10.1021/nl501425n} {\bibfield
  {journal} {\bibinfo  {journal} {Nano Lett.}\ }\textbf {\bibinfo {volume}
  {14}},\ \bibinfo {pages} {4011} (\bibinfo {year} {2014})}\BibitemShut
  {NoStop}%
\bibitem [{\citenamefont {Chen}\ \emph {et~al.}(2011)\citenamefont {Chen},
  \citenamefont {Li}, \citenamefont {Cullen}, \citenamefont {Williams},\ and\
  \citenamefont {Fuhrer}}]{Fuhrer}%
  \BibitemOpen
  \bibfield  {author} {\bibinfo {author} {\bibfnamefont {J.-H.}\ \bibnamefont
  {Chen}}, \bibinfo {author} {\bibfnamefont {L.}~\bibnamefont {Li}}, \bibinfo
  {author} {\bibfnamefont {W.~G.}\ \bibnamefont {Cullen}}, \bibinfo {author}
  {\bibfnamefont {E.~D.}\ \bibnamefont {Williams}}, \ and\ \bibinfo {author}
  {\bibfnamefont {M.~S.}\ \bibnamefont {Fuhrer}},\ }\href {\doibase
  10.1038/nphys1962} {\bibfield  {journal} {\bibinfo  {journal} {Nat. Phys.}\
  }\textbf {\bibinfo {volume} {7}},\ \bibinfo {pages} {535} (\bibinfo {year}
  {2011})}\BibitemShut {NoStop}%
\bibitem [{\citenamefont {Miranda}\ \emph {et~al.}(2014)\citenamefont
  {Miranda}, \citenamefont {Dias~da Silva},\ and\ \citenamefont
  {Lewenkopf}}]{PhysRevB.90.201101}%
  \BibitemOpen
  \bibfield  {author} {\bibinfo {author} {\bibfnamefont {V.~G.}\ \bibnamefont
  {Miranda}}, \bibinfo {author} {\bibfnamefont {L.~G. G.~V.}\ \bibnamefont
  {Dias~da Silva}}, \ and\ \bibinfo {author} {\bibfnamefont {C.~H.}\
  \bibnamefont {Lewenkopf}},\ }\href {\doibase 10.1103/PhysRevB.90.201101}
  {\bibfield  {journal} {\bibinfo  {journal} {Phys. Rev. B}\ }\textbf {\bibinfo
  {volume} {90}},\ \bibinfo {pages} {201101} (\bibinfo {year}
  {2014})}\BibitemShut {NoStop}%
\bibitem [{\citenamefont {B\"usser}\ \emph {et~al.}(2013)\citenamefont
  {B\"usser}, \citenamefont {Martins},\ and\ \citenamefont
  {Feiguin}}]{Busser2013}%
  \BibitemOpen
  \bibfield  {author} {\bibinfo {author} {\bibfnamefont {C.~A.}\ \bibnamefont
  {B\"usser}}, \bibinfo {author} {\bibfnamefont {G.~B.}\ \bibnamefont
  {Martins}}, \ and\ \bibinfo {author} {\bibfnamefont {A.~E.}\ \bibnamefont
  {Feiguin}},\ }\href {\doibase 10.1103/PhysRevB.88.245113} {\bibfield
  {journal} {\bibinfo  {journal} {Phys. Rev. B}\ }\textbf {\bibinfo {volume}
  {88}},\ \bibinfo {pages} {245113} (\bibinfo {year} {2013})}\BibitemShut
  {NoStop}%
\bibitem [{\citenamefont {Krychowski}\ \emph {et~al.}(2014)\citenamefont
  {Krychowski}, \citenamefont {Kaczkowski},\ and\ \citenamefont
  {Lipinski}}]{PhysRevB.89.035424}%
  \BibitemOpen
  \bibfield  {author} {\bibinfo {author} {\bibfnamefont {D.}~\bibnamefont
  {Krychowski}}, \bibinfo {author} {\bibfnamefont {J.}~\bibnamefont
  {Kaczkowski}}, \ and\ \bibinfo {author} {\bibfnamefont {S.}~\bibnamefont
  {Lipinski}},\ }\href {\doibase 10.1103/PhysRevB.89.035424} {\bibfield
  {journal} {\bibinfo  {journal} {Phys. Rev. B}\ }\textbf {\bibinfo {volume}
  {89}},\ \bibinfo {pages} {035424} (\bibinfo {year} {2014})}\BibitemShut
  {NoStop}%
\bibitem [{\citenamefont {Li}\ \emph {et~al.}(2017)\citenamefont {Li},
  \citenamefont {Ngo}, \citenamefont {DiLullo}, \citenamefont {Latt},
  \citenamefont {Kersell}, \citenamefont {Fisher}, \citenamefont {Zapol},\ and\
  \citenamefont {Ulloa}}]{Li2017}%
  \BibitemOpen
  \bibfield  {author} {\bibinfo {author} {\bibfnamefont {Y.}~\bibnamefont
  {Li}}, \bibinfo {author} {\bibfnamefont {A.~T.}\ \bibnamefont {Ngo}},
  \bibinfo {author} {\bibfnamefont {A.}~\bibnamefont {DiLullo}}, \bibinfo
  {author} {\bibfnamefont {K.~Z.}\ \bibnamefont {Latt}}, \bibinfo {author}
  {\bibfnamefont {H.}~\bibnamefont {Kersell}}, \bibinfo {author} {\bibfnamefont
  {B.}~\bibnamefont {Fisher}}, \bibinfo {author} {\bibfnamefont
  {P.}~\bibnamefont {Zapol}}, \ and\ \bibinfo {author} {\bibfnamefont {S.-W.}\
  \bibnamefont {Ulloa}, \bibfnamefont {Sergio E.and~Hla}},\ }\href {\doibase
  10.1038/s41467-017-00881-1} {\bibfield  {journal} {\bibinfo  {journal} {Nat.
  Commun.}\ }\textbf {\bibinfo {volume} {8}},\ \bibinfo {pages} {946} (\bibinfo
  {year} {2017})}\BibitemShut {NoStop}%
\bibitem [{\citenamefont {Diniz}\ \emph {et~al.}(2018)\citenamefont {Diniz},
  \citenamefont {Luiz}, \citenamefont {Latg\'e},\ and\ \citenamefont
  {Vernek}}]{PhysRevB.97.115444}%
  \BibitemOpen
  \bibfield  {author} {\bibinfo {author} {\bibfnamefont {G.~S.}\ \bibnamefont
  {Diniz}}, \bibinfo {author} {\bibfnamefont {G.~I.}\ \bibnamefont {Luiz}},
  \bibinfo {author} {\bibfnamefont {A.}~\bibnamefont {Latg\'e}}, \ and\
  \bibinfo {author} {\bibfnamefont {E.}~\bibnamefont {Vernek}},\ }\href
  {\doibase 10.1103/PhysRevB.97.115444} {\bibfield  {journal} {\bibinfo
  {journal} {Phys. Rev. B}\ }\textbf {\bibinfo {volume} {97}},\ \bibinfo
  {pages} {115444} (\bibinfo {year} {2018})}\BibitemShut {NoStop}%
\bibitem [{\citenamefont {Wakabayashi}\ \emph {et~al.}(2009)\citenamefont
  {Wakabayashi}, \citenamefont {Takane}, \citenamefont {Yamamoto},\ and\
  \citenamefont {Sigrist}}]{Wakabayashi2009}%
  \BibitemOpen
  \bibfield  {author} {\bibinfo {author} {\bibfnamefont {K.}~\bibnamefont
  {Wakabayashi}}, \bibinfo {author} {\bibfnamefont {Y.}~\bibnamefont {Takane}},
  \bibinfo {author} {\bibfnamefont {M.}~\bibnamefont {Yamamoto}}, \ and\
  \bibinfo {author} {\bibfnamefont {M.}~\bibnamefont {Sigrist}},\ }\href
  {\doibase 10.1088/1367-2630/11/9/095016} {\bibfield  {journal} {\bibinfo
  {journal} {New J. Phys.}\ }\textbf {\bibinfo {volume} {11}},\ \bibinfo
  {pages} {095016} (\bibinfo {year} {2009})}\BibitemShut {NoStop}%
\bibitem [{\citenamefont {Anderson}(1970)}]{Anderson1970}%
  \BibitemOpen
  \bibfield  {author} {\bibinfo {author} {\bibfnamefont {P.~W.}\ \bibnamefont
  {Anderson}},\ }\href {\doibase 10.1088/0022-3719/3/12/008} {\bibfield
  {journal} {\bibinfo  {journal} {J. Phys. C: Sol. St. Phys.}\ }\textbf
  {\bibinfo {volume} {3}},\ \bibinfo {pages} {2436} (\bibinfo {year}
  {1970})}\BibitemShut {NoStop}%
\bibitem [{\citenamefont {Cornaglia}\ and\ \citenamefont
  {Grempel}(2005)}]{Cornaglia2005}%
  \BibitemOpen
  \bibfield  {author} {\bibinfo {author} {\bibfnamefont {P.~S.}\ \bibnamefont
  {Cornaglia}}\ and\ \bibinfo {author} {\bibfnamefont {D.~R.}\ \bibnamefont
  {Grempel}},\ }\href {\doibase 10.1103/PhysRevB.71.075305} {\bibfield
  {journal} {\bibinfo  {journal} {Phys. Rev. B}\ }\textbf {\bibinfo {volume}
  {71}},\ \bibinfo {pages} {075305} (\bibinfo {year} {2005})}\BibitemShut
  {NoStop}%
\bibitem [{Lju()}]{Ljubljana}%
  \BibitemOpen
  \href@noop {} {}\bibinfo {note} {R. \v{Z}itko, NRG Ljubljana - open source
  NRG code available at http://nrgljubljana.ijs.si.}\BibitemShut {Stop}%
\bibitem [{TK1()}]{TK1}%
  \BibitemOpen
  \href@noop {} {}\bibinfo {note} {As will be more clearly explained later,
  what we call $T_K$ in this section will become $T_{K1}$, once we introduce a
  finite $\Gamma_{\rm M}$ to the calculations. However, in this subsection, to
  avoid confusion, we will refer to the ($\Gamma_{\rm M}=0$) PMS Kondo
  temperature simply as $T_K$.}\BibitemShut {Stop}%
\bibitem [{\citenamefont {Schrieffer}\ and\ \citenamefont
  {Wolff}(1966)}]{Schrieffer1966}%
  \BibitemOpen
  \bibfield  {author} {\bibinfo {author} {\bibfnamefont {J.~R.}\ \bibnamefont
  {Schrieffer}}\ and\ \bibinfo {author} {\bibfnamefont {P.~A.}\ \bibnamefont
  {Wolff}},\ }\href {\doibase 10.1103/PhysRev.149.491} {\bibfield  {journal}
  {\bibinfo  {journal} {Phys. Rev.}\ }\textbf {\bibinfo {volume} {149}},\
  \bibinfo {pages} {491} (\bibinfo {year} {1966})}\BibitemShut {NoStop}%
\bibitem [{not()}]{note0}%
  \BibitemOpen
  \href@noop {} {}\bibinfo {note} {Equivalently, we could also choose to look
  for a critical $\Delta_c$ for a given $J_S^{(0)}$.}\BibitemShut {Stop}%
\bibitem [{rho()}]{rho0}%
  \BibitemOpen
  \href@noop {} {}\bibinfo {note} {We have used the fact that $\rho_S(D)=\rho_0
  = \nicefrac{1}{2\sqrt{D^2-\Delta^2}}$, which, for $\Delta=0$, results in
  $\rho_0 = \nicefrac{1}{2D}$.}\BibitemShut {Stop}%
\bibitem [{\citenamefont {Zubarev}(1960)}]{Zubarev_1960}%
  \BibitemOpen
  \bibfield  {author} {\bibinfo {author} {\bibfnamefont {D.~N.}\ \bibnamefont
  {Zubarev}},\ }\href {\doibase 10.1070/pu1960v003n03abeh003275} {\bibfield
  {journal} {\bibinfo  {journal} {Phys.-Uspekhi}\ }\textbf {\bibinfo {volume}
  {3}},\ \bibinfo {pages} {320} (\bibinfo {year} {1960})}\BibitemShut {NoStop}%
\bibitem [{\citenamefont {Jefferson}(1977)}]{Jefferson_1977}%
  \BibitemOpen
  \bibfield  {author} {\bibinfo {author} {\bibfnamefont {J.~H.}\ \bibnamefont
  {Jefferson}},\ }\href {\doibase 10.1088/0022-3719/10/18/023} {\bibfield
  {journal} {\bibinfo  {journal} {J. Phys. C: Sol. St. Phys.}\ }\textbf
  {\bibinfo {volume} {10}},\ \bibinfo {pages} {3589} (\bibinfo {year}
  {1977})}\BibitemShut {NoStop}%
\bibitem [{\citenamefont {Haldane}(1978)}]{Haldane1978}%
  \BibitemOpen
  \bibfield  {author} {\bibinfo {author} {\bibfnamefont {F.~D.~M.}\
  \bibnamefont {Haldane}},\ }\href {\doibase 10.1103/PhysRevLett.40.416}
  {\bibfield  {journal} {\bibinfo  {journal} {Phys. Rev. Lett.}\ }\textbf
  {\bibinfo {volume} {40}},\ \bibinfo {pages} {416} (\bibinfo {year}
  {1978})}\BibitemShut {NoStop}%
\bibitem [{\citenamefont {Cheng}\ and\ \citenamefont
  {Ingersent}(2013)}]{Cheng2013}%
  \BibitemOpen
  \bibfield  {author} {\bibinfo {author} {\bibfnamefont {M.}~\bibnamefont
  {Cheng}}\ and\ \bibinfo {author} {\bibfnamefont {K.}~\bibnamefont
  {Ingersent}},\ }\href {\doibase 10.1103/PhysRevB.87.075145} {\bibfield
  {journal} {\bibinfo  {journal} {Phys. Rev. B}\ }\textbf {\bibinfo {volume}
  {87}},\ \bibinfo {pages} {075145} (\bibinfo {year} {2013})}\BibitemShut
  {NoStop}%
\bibitem [{\citenamefont {Cheng}\ \emph {et~al.}(2017)\citenamefont {Cheng},
  \citenamefont {Chowdhury}, \citenamefont {Mohammed},\ and\ \citenamefont
  {Ingersent}}]{Kevin2017}%
  \BibitemOpen
  \bibfield  {author} {\bibinfo {author} {\bibfnamefont {M.}~\bibnamefont
  {Cheng}}, \bibinfo {author} {\bibfnamefont {T.}~\bibnamefont {Chowdhury}},
  \bibinfo {author} {\bibfnamefont {A.}~\bibnamefont {Mohammed}}, \ and\
  \bibinfo {author} {\bibfnamefont {K.}~\bibnamefont {Ingersent}},\ }\href
  {\doibase 10.1103/PhysRevB.96.045103} {\bibfield  {journal} {\bibinfo
  {journal} {Phys. Rev. B}\ }\textbf {\bibinfo {volume} {96}},\ \bibinfo
  {pages} {045103} (\bibinfo {year} {2017})}\BibitemShut {NoStop}%
\bibitem [{\citenamefont {Massidda}\ \emph {et~al.}(1990)\citenamefont
  {Massidda}, \citenamefont {Continenza}, \citenamefont {Freeman},
  \citenamefont {de~Pascale}, \citenamefont {Meloni},\ and\ \citenamefont
  {Serra}}]{Massidda1990}%
  \BibitemOpen
  \bibfield  {author} {\bibinfo {author} {\bibfnamefont {S.}~\bibnamefont
  {Massidda}}, \bibinfo {author} {\bibfnamefont {A.}~\bibnamefont
  {Continenza}}, \bibinfo {author} {\bibfnamefont {A.~J.}\ \bibnamefont
  {Freeman}}, \bibinfo {author} {\bibfnamefont {T.~M.}\ \bibnamefont
  {de~Pascale}}, \bibinfo {author} {\bibfnamefont {F.}~\bibnamefont {Meloni}},
  \ and\ \bibinfo {author} {\bibfnamefont {M.}~\bibnamefont {Serra}},\ }\href
  {\doibase 10.1103/PhysRevB.41.12079} {\bibfield  {journal} {\bibinfo
  {journal} {Phys. Rev. B}\ }\textbf {\bibinfo {volume} {41}},\ \bibinfo
  {pages} {12079} (\bibinfo {year} {1990})}\BibitemShut {NoStop}%
\bibitem [{\citenamefont {Heeger}\ \emph {et~al.}(1988)\citenamefont {Heeger},
  \citenamefont {Kivelson}, \citenamefont {Schrieffer},\ and\ \citenamefont
  {Su}}]{Heeger1988}%
  \BibitemOpen
  \bibfield  {author} {\bibinfo {author} {\bibfnamefont {A.~J.}\ \bibnamefont
  {Heeger}}, \bibinfo {author} {\bibfnamefont {S.}~\bibnamefont {Kivelson}},
  \bibinfo {author} {\bibfnamefont {J.~R.}\ \bibnamefont {Schrieffer}}, \ and\
  \bibinfo {author} {\bibfnamefont {W.~P.}\ \bibnamefont {Su}},\ }\href
  {\doibase 10.1103/RevModPhys.60.781} {\bibfield  {journal} {\bibinfo
  {journal} {Rev. Mod. Phys.}\ }\textbf {\bibinfo {volume} {60}},\ \bibinfo
  {pages} {781} (\bibinfo {year} {1988})}\BibitemShut {NoStop}%
\bibitem [{\citenamefont {Borghardt}\ \emph {et~al.}(2017)\citenamefont
  {Borghardt}, \citenamefont {Tu}, \citenamefont {Winkler}, \citenamefont
  {Schubert}, \citenamefont {Zander}, \citenamefont {Leosson},\ and\
  \citenamefont {Kardyna{\l}}}]{Borghardt2017}%
  \BibitemOpen
  \bibfield  {author} {\bibinfo {author} {\bibfnamefont {S.}~\bibnamefont
  {Borghardt}}, \bibinfo {author} {\bibfnamefont {J.~S.}\ \bibnamefont {Tu}},
  \bibinfo {author} {\bibfnamefont {F.}~\bibnamefont {Winkler}}, \bibinfo
  {author} {\bibfnamefont {J.}~\bibnamefont {Schubert}}, \bibinfo {author}
  {\bibfnamefont {W.}~\bibnamefont {Zander}}, \bibinfo {author} {\bibfnamefont
  {K.}~\bibnamefont {Leosson}}, \ and\ \bibinfo {author} {\bibfnamefont
  {B.~E.}\ \bibnamefont {Kardyna{\l}}},\ }\href {\doibase
  10.1103/PhysRevMaterials.1.054001} {\bibfield  {journal} {\bibinfo  {journal}
  {Phys. Rev. Materials}\ }\textbf {\bibinfo {volume} {1}},\ \bibinfo {pages}
  {1} (\bibinfo {year} {2017})}\BibitemShut {NoStop}%
\bibitem [{\citenamefont {Lenz}\ \emph {et~al.}(2013)\citenamefont {Lenz},
  \citenamefont {Urban},\ and\ \citenamefont {Bercioux}}]{Lenz2013}%
  \BibitemOpen
  \bibfield  {author} {\bibinfo {author} {\bibfnamefont {L.}~\bibnamefont
  {Lenz}}, \bibinfo {author} {\bibfnamefont {D.~F.}\ \bibnamefont {Urban}}, \
  and\ \bibinfo {author} {\bibfnamefont {D.}~\bibnamefont {Bercioux}},\ }\href
  {\doibase 10.1140/epjb/e2013-40760-4} {\bibfield  {journal} {\bibinfo
  {journal} {Eur. Phys. J. B}\ }\textbf {\bibinfo {volume} {86}},\ \bibinfo
  {pages} {502} (\bibinfo {year} {2013})}\BibitemShut {NoStop}%
\bibitem [{Dim()}]{Dimmers}%
  \BibitemOpen
  \href@noop {} {}\bibinfo {note} {Depending on the number of dimmers along the
  transverse direction, the AGNR can be metallic or an \emph{intrinsic}
  semiconductor. In what follows, we will explore the extrinsic tunable induced
  gap in metallic AGNR by means of an external agent.}\BibitemShut {Stop}%
\bibitem [{\citenamefont {Pereira}\ \emph {et~al.}(2009)\citenamefont
  {Pereira}, \citenamefont {Castro~Neto},\ and\ \citenamefont
  {Peres}}]{PhysRevB.80.045401}%
  \BibitemOpen
  \bibfield  {author} {\bibinfo {author} {\bibfnamefont {V.~M.}\ \bibnamefont
  {Pereira}}, \bibinfo {author} {\bibfnamefont {A.~H.}\ \bibnamefont
  {Castro~Neto}}, \ and\ \bibinfo {author} {\bibfnamefont {N.~M.~R.}\
  \bibnamefont {Peres}},\ }\href {\doibase 10.1103/PhysRevB.80.045401}
  {\bibfield  {journal} {\bibinfo  {journal} {Phys. Rev. B}\ }\textbf {\bibinfo
  {volume} {80}},\ \bibinfo {pages} {045401} (\bibinfo {year}
  {2009})}\BibitemShut {NoStop}%
\bibitem [{\citenamefont {Castro~Neto}\ \emph {et~al.}(2009)\citenamefont
  {Castro~Neto}, \citenamefont {Guinea}, \citenamefont {Peres}, \citenamefont
  {Novoselov},\ and\ \citenamefont {Geim}}]{RevModPhys.81.109}%
  \BibitemOpen
  \bibfield  {author} {\bibinfo {author} {\bibfnamefont {A.~H.}\ \bibnamefont
  {Castro~Neto}}, \bibinfo {author} {\bibfnamefont {F.}~\bibnamefont {Guinea}},
  \bibinfo {author} {\bibfnamefont {N.~M.~R.}\ \bibnamefont {Peres}}, \bibinfo
  {author} {\bibfnamefont {K.~S.}\ \bibnamefont {Novoselov}}, \ and\ \bibinfo
  {author} {\bibfnamefont {A.~K.}\ \bibnamefont {Geim}},\ }\href {\doibase
  10.1103/RevModPhys.81.109} {\bibfield  {journal} {\bibinfo  {journal} {Rev.
  Mod. Phys.}\ }\textbf {\bibinfo {volume} {81}},\ \bibinfo {pages} {109}
  (\bibinfo {year} {2009})}\BibitemShut {NoStop}%
\bibitem [{\citenamefont {Kane}\ and\ \citenamefont {Mele}(2005)}]{Kane2005}%
  \BibitemOpen
  \bibfield  {author} {\bibinfo {author} {\bibfnamefont {C.~L.}\ \bibnamefont
  {Kane}}\ and\ \bibinfo {author} {\bibfnamefont {E.~J.}\ \bibnamefont
  {Mele}},\ }\href {\doibase 10.1103/PhysRevLett.95.226801} {\bibfield
  {journal} {\bibinfo  {journal} {Phys. Rev. Lett.}\ }\textbf {\bibinfo
  {volume} {95}},\ \bibinfo {pages} {226801} (\bibinfo {year}
  {2005})}\BibitemShut {NoStop}%
\bibitem [{\citenamefont {Zarea}\ and\ \citenamefont
  {Sandler}(2009)}]{PhysRevB.79.165442}%
  \BibitemOpen
  \bibfield  {author} {\bibinfo {author} {\bibfnamefont {M.}~\bibnamefont
  {Zarea}}\ and\ \bibinfo {author} {\bibfnamefont {N.}~\bibnamefont
  {Sandler}},\ }\href {\doibase 10.1103/PhysRevB.79.165442} {\bibfield
  {journal} {\bibinfo  {journal} {Phys. Rev. B}\ }\textbf {\bibinfo {volume}
  {79}},\ \bibinfo {pages} {165442} (\bibinfo {year} {2009})}\BibitemShut
  {NoStop}%
\bibitem [{\citenamefont {Nardelli}(1999)}]{Nardelli1999}%
  \BibitemOpen
  \bibfield  {author} {\bibinfo {author} {\bibfnamefont {M.~B.}\ \bibnamefont
  {Nardelli}},\ }\href {\doibase 10.1103/PhysRevB.60.7828} {\bibfield
  {journal} {\bibinfo  {journal} {Phys. Rev. B}\ }\textbf {\bibinfo {volume}
  {60}},\ \bibinfo {pages} {7828} (\bibinfo {year} {1999})}\BibitemShut
  {NoStop}%
\bibitem [{\citenamefont {Sancho}\ \emph {et~al.}(1984)\citenamefont {Sancho},
  \citenamefont {Sancho},\ and\ \citenamefont {Rubio}}]{Sancho1984}%
  \BibitemOpen
  \bibfield  {author} {\bibinfo {author} {\bibfnamefont {M.~P.~L.}\
  \bibnamefont {Sancho}}, \bibinfo {author} {\bibfnamefont {J.~M.~L.}\
  \bibnamefont {Sancho}}, \ and\ \bibinfo {author} {\bibfnamefont
  {J.}~\bibnamefont {Rubio}},\ }\href@noop {} {\bibfield  {journal} {\bibinfo
  {journal} {J. Phys. F: Met. Phys.}\ }\textbf {\bibinfo {volume} {14}},\
  \bibinfo {pages} {1205} (\bibinfo {year} {1984})}\BibitemShut {NoStop}%
\bibitem [{AGN()}]{AGNR+tip}%
  \BibitemOpen
  \href@noop {} {}\bibinfo {note} {Thus, the hybridization function just
  described is equivalent to the $\Gamma_0=\Gamma_{\rm S}+\Gamma_{\rm M}$
  hybridization function defined in Sec.~\ref{sec:model}}\BibitemShut {NoStop}%
\bibitem [{top()}]{topsite}%
  \BibitemOpen
  \href@noop {} {}\bibinfo {note} {Co adatoms, deposited on monolayer graphene
  (deposited, in turn, over a SiC (0001) substrate), have shown to be most
  favorable in a top-site configuration (see Fig.~\ref{fig1}), as
  experimentally reported by Eelbo \emph{et al}., Phys. Rev. Lett. 110, 136804
  (2013).}\BibitemShut {Stop}%
\bibitem [{\citenamefont {Demchenko}\ \emph {et~al.}(2004)\citenamefont
  {Demchenko}, \citenamefont {Joura},\ and\ \citenamefont
  {Freericks}}]{PhysRevLett.92.216401}%
  \BibitemOpen
  \bibfield  {author} {\bibinfo {author} {\bibfnamefont {D.~O.}\ \bibnamefont
  {Demchenko}}, \bibinfo {author} {\bibfnamefont {A.~V.}\ \bibnamefont
  {Joura}}, \ and\ \bibinfo {author} {\bibfnamefont {J.~K.}\ \bibnamefont
  {Freericks}},\ }\href {\doibase 10.1103/PhysRevLett.92.216401} {\bibfield
  {journal} {\bibinfo  {journal} {Phys. Rev. Lett.}\ }\textbf {\bibinfo
  {volume} {92}},\ \bibinfo {pages} {216401} (\bibinfo {year}
  {2004})}\BibitemShut {NoStop}%
\bibitem [{\citenamefont {Rashba}(2009)}]{PhysRevB.79.161409}%
  \BibitemOpen
  \bibfield  {author} {\bibinfo {author} {\bibfnamefont {E.~I.}\ \bibnamefont
  {Rashba}},\ }\href {\doibase 10.1103/PhysRevB.79.161409} {\bibfield
  {journal} {\bibinfo  {journal} {Phys. Rev. B}\ }\textbf {\bibinfo {volume}
  {79}},\ \bibinfo {pages} {161409} (\bibinfo {year} {2009})}\BibitemShut
  {NoStop}%
\bibitem [{\citenamefont {Dedkov}\ \emph {et~al.}(2008)\citenamefont {Dedkov},
  \citenamefont {Fonin}, \citenamefont {R\"udiger},\ and\ \citenamefont
  {Laubschat}}]{PhysRevLett.100.107602}%
  \BibitemOpen
  \bibfield  {author} {\bibinfo {author} {\bibfnamefont {Y.~S.}\ \bibnamefont
  {Dedkov}}, \bibinfo {author} {\bibfnamefont {M.}~\bibnamefont {Fonin}},
  \bibinfo {author} {\bibfnamefont {U.}~\bibnamefont {R\"udiger}}, \ and\
  \bibinfo {author} {\bibfnamefont {C.}~\bibnamefont {Laubschat}},\ }\href
  {\doibase 10.1103/PhysRevLett.100.107602} {\bibfield  {journal} {\bibinfo
  {journal} {Phys. Rev. Lett.}\ }\textbf {\bibinfo {volume} {100}},\ \bibinfo
  {pages} {107602} (\bibinfo {year} {2008})}\BibitemShut {NoStop}%
\bibitem [{\citenamefont {Varykhalov}\ \emph {et~al.}(2012)\citenamefont
  {Varykhalov}, \citenamefont {Marchenko}, \citenamefont {Scholz},
  \citenamefont {Rienks}, \citenamefont {Kim}, \citenamefont {Bihlmayer},
  \citenamefont {S\'anchez-Barriga},\ and\ \citenamefont
  {Rader}}]{PhysRevLett.108.066804}%
  \BibitemOpen
  \bibfield  {author} {\bibinfo {author} {\bibfnamefont {A.}~\bibnamefont
  {Varykhalov}}, \bibinfo {author} {\bibfnamefont {D.}~\bibnamefont
  {Marchenko}}, \bibinfo {author} {\bibfnamefont {M.~R.}\ \bibnamefont
  {Scholz}}, \bibinfo {author} {\bibfnamefont {E.~D.~L.}\ \bibnamefont
  {Rienks}}, \bibinfo {author} {\bibfnamefont {T.~K.}\ \bibnamefont {Kim}},
  \bibinfo {author} {\bibfnamefont {G.}~\bibnamefont {Bihlmayer}}, \bibinfo
  {author} {\bibfnamefont {J.}~\bibnamefont {S\'anchez-Barriga}}, \ and\
  \bibinfo {author} {\bibfnamefont {O.}~\bibnamefont {Rader}},\ }\href
  {\doibase 10.1103/PhysRevLett.108.066804} {\bibfield  {journal} {\bibinfo
  {journal} {Phys. Rev. Lett.}\ }\textbf {\bibinfo {volume} {108}},\ \bibinfo
  {pages} {066804} (\bibinfo {year} {2012})}\BibitemShut {NoStop}%
\bibitem [{\citenamefont {Marchenko}\ \emph {et~al.}(2012)\citenamefont
  {Marchenko}, \citenamefont {Varykhalov}, \citenamefont {Scholz},
  \citenamefont {Bihlmayer}, \citenamefont {Rashba}, \citenamefont {Rybkin},
  \citenamefont {Shikin},\ and\ \citenamefont {Rader}}]{Marchenko2012}%
  \BibitemOpen
  \bibfield  {author} {\bibinfo {author} {\bibfnamefont {D.}~\bibnamefont
  {Marchenko}}, \bibinfo {author} {\bibfnamefont {A.}~\bibnamefont
  {Varykhalov}}, \bibinfo {author} {\bibfnamefont {M.~R.}\ \bibnamefont
  {Scholz}}, \bibinfo {author} {\bibfnamefont {G.}~\bibnamefont {Bihlmayer}},
  \bibinfo {author} {\bibfnamefont {E.~I.}\ \bibnamefont {Rashba}}, \bibinfo
  {author} {\bibfnamefont {A.}~\bibnamefont {Rybkin}}, \bibinfo {author}
  {\bibfnamefont {A.~M.}\ \bibnamefont {Shikin}}, \ and\ \bibinfo {author}
  {\bibfnamefont {O.}~\bibnamefont {Rader}},\ }\href {\doibase
  10.1038/ncomms2227} {\bibfield  {journal} {\bibinfo  {journal} {Nat.
  Commun.}\ }\textbf {\bibinfo {volume} {3}},\ \bibinfo {pages} {1232}
  (\bibinfo {year} {2012})}\BibitemShut {NoStop}%
\bibitem [{van()}]{van-Hove}%
  \BibitemOpen
  \href@noop {} {}\bibinfo {note} {The van-Hove singularities are fixed for a
  given $\lambda_{R}$, however, as $\lambda_{R}$ is varied (for the range of
  $\lambda_{R}$ considered), the first peak is strongly modified, as one can
  note in Fig.\ref{delta}.}\BibitemShut {Stop}%
\bibitem [{hyb()}]{hybrid}%
  \BibitemOpen
  \href@noop {} {}\bibinfo {note} {Note that the AGNR used for these
  calculations had a width ($W\approx11.56$nm) different from the one used for
  the calculations in Fig.~\ref{fig3-AGNR} ($W\approx5.65$nm). This resulted in
  a $\Gamma_{\rm S}$ (not shown) considerably smaller than the one obtained for
  the AGNR in Fig.~\ref{fig3-AGNR}, thus a smaller $T_{K1}$ than the one
  expected from just a decrease in $U$.}\BibitemShut {Stop}%
\bibitem [{\citenamefont {Pustilnik}\ and\ \citenamefont
  {Glazman}(2001)}]{Pustilnik2001}%
  \BibitemOpen
  \bibfield  {author} {\bibinfo {author} {\bibfnamefont {M.}~\bibnamefont
  {Pustilnik}}\ and\ \bibinfo {author} {\bibfnamefont {L.~I.}\ \bibnamefont
  {Glazman}},\ }\href {\doibase 10.1103/PhysRevLett.87.216601} {\bibfield
  {journal} {\bibinfo  {journal} {Phys. Rev. Lett.}\ }\textbf {\bibinfo
  {volume} {87}},\ \bibinfo {pages} {216601} (\bibinfo {year}
  {2001})}\BibitemShut {NoStop}%
\bibitem [{\citenamefont {Hofstetter}\ and\ \citenamefont
  {Zarand}(2004)}]{Hofstetter2004}%
  \BibitemOpen
  \bibfield  {author} {\bibinfo {author} {\bibfnamefont {W.}~\bibnamefont
  {Hofstetter}}\ and\ \bibinfo {author} {\bibfnamefont {G.}~\bibnamefont
  {Zarand}},\ }\href {\doibase 10.1103/PhysRevB.69.235301} {\bibfield
  {journal} {\bibinfo  {journal} {Phys. Rev. B}\ }\textbf {\bibinfo {volume}
  {69}},\ \bibinfo {pages} {235301} (\bibinfo {year} {2004})}\BibitemShut
  {NoStop}%
\bibitem [{\citenamefont {\v{Z}itko}(2010)}]{Zitko2010}%
  \BibitemOpen
  \bibfield  {author} {\bibinfo {author} {\bibfnamefont {R.}~\bibnamefont
  {\v{Z}itko}},\ }\href {\doibase 10.1103/PhysRevB.81.115316} {\bibfield
  {journal} {\bibinfo  {journal} {Phys. Rev. B}\ }\textbf {\bibinfo {volume}
  {81}},\ \bibinfo {pages} {115316} (\bibinfo {year} {2010})}\BibitemShut
  {NoStop}%
\bibitem [{\citenamefont {van~der Wiel}\ \emph {et~al.}(2002)\citenamefont
  {van~der Wiel}, \citenamefont {De~Franceschi}, \citenamefont {Elzerman},
  \citenamefont {Tarucha}, \citenamefont {Kouwenhoven}, \citenamefont
  {Motohisa}, \citenamefont {Nakajima},\ and\ \citenamefont
  {Fukui}}]{Wiel2002}%
  \BibitemOpen
  \bibfield  {author} {\bibinfo {author} {\bibfnamefont {W.~G.}\ \bibnamefont
  {van~der Wiel}}, \bibinfo {author} {\bibfnamefont {S.}~\bibnamefont
  {De~Franceschi}}, \bibinfo {author} {\bibfnamefont {J.~M.}\ \bibnamefont
  {Elzerman}}, \bibinfo {author} {\bibfnamefont {S.}~\bibnamefont {Tarucha}},
  \bibinfo {author} {\bibfnamefont {L.~P.}\ \bibnamefont {Kouwenhoven}},
  \bibinfo {author} {\bibfnamefont {J.}~\bibnamefont {Motohisa}}, \bibinfo
  {author} {\bibfnamefont {F.}~\bibnamefont {Nakajima}}, \ and\ \bibinfo
  {author} {\bibfnamefont {T.}~\bibnamefont {Fukui}},\ }\href {\doibase
  10.1103/PhysRevLett.88.126803} {\bibfield  {journal} {\bibinfo  {journal}
  {Phys. Rev. Lett.}\ }\textbf {\bibinfo {volume} {88}},\ \bibinfo {pages}
  {126803} (\bibinfo {year} {2002})}\BibitemShut {NoStop}%
\bibitem [{\citenamefont {Granger}\ \emph {et~al.}(2005)\citenamefont
  {Granger}, \citenamefont {Kastner}, \citenamefont {Radu}, \citenamefont
  {Hanson},\ and\ \citenamefont {Gossard}}]{Granger2005}%
  \BibitemOpen
  \bibfield  {author} {\bibinfo {author} {\bibfnamefont {G.}~\bibnamefont
  {Granger}}, \bibinfo {author} {\bibfnamefont {M.~A.}\ \bibnamefont
  {Kastner}}, \bibinfo {author} {\bibfnamefont {I.}~\bibnamefont {Radu}},
  \bibinfo {author} {\bibfnamefont {M.~P.}\ \bibnamefont {Hanson}}, \ and\
  \bibinfo {author} {\bibfnamefont {A.~C.}\ \bibnamefont {Gossard}},\ }\href
  {\doibase 10.1103/PhysRevB.72.165309} {\bibfield  {journal} {\bibinfo
  {journal} {Phys. Rev. B}\ }\textbf {\bibinfo {volume} {72}},\ \bibinfo
  {pages} {165309} (\bibinfo {year} {2005})}\BibitemShut {NoStop}%
\bibitem [{\citenamefont {Sasaki}\ \emph {et~al.}(2009)\citenamefont {Sasaki},
  \citenamefont {Tamura}, \citenamefont {Akazaki},\ and\ \citenamefont
  {Fujisawa}}]{Sasaki2009}%
  \BibitemOpen
  \bibfield  {author} {\bibinfo {author} {\bibfnamefont {S.}~\bibnamefont
  {Sasaki}}, \bibinfo {author} {\bibfnamefont {H.}~\bibnamefont {Tamura}},
  \bibinfo {author} {\bibfnamefont {T.}~\bibnamefont {Akazaki}}, \ and\
  \bibinfo {author} {\bibfnamefont {T.}~\bibnamefont {Fujisawa}},\ }\href
  {\doibase 10.1103/PhysRevLett.103.266806} {\bibfield  {journal} {\bibinfo
  {journal} {Phys. Rev. Lett.}\ }\textbf {\bibinfo {volume} {103}},\ \bibinfo
  {pages} {266806} (\bibinfo {year} {2009})}\BibitemShut {NoStop}%
\bibitem [{\citenamefont {Sasaki}\ \emph {et~al.}(2000)\citenamefont {Sasaki},
  \citenamefont {De~Franceschi}, \citenamefont {Elzerman}, \citenamefont
  {van~der Wiel}, \citenamefont {Eto}, \citenamefont {Tarucha},\ and\
  \citenamefont {Kouwenhoven}}]{Sasaki2000}%
  \BibitemOpen
  \bibfield  {author} {\bibinfo {author} {\bibfnamefont {S.}~\bibnamefont
  {Sasaki}}, \bibinfo {author} {\bibfnamefont {S.}~\bibnamefont
  {De~Franceschi}}, \bibinfo {author} {\bibfnamefont {J.~M.}\ \bibnamefont
  {Elzerman}}, \bibinfo {author} {\bibfnamefont {W.~G.}\ \bibnamefont {van~der
  Wiel}}, \bibinfo {author} {\bibfnamefont {M.}~\bibnamefont {Eto}}, \bibinfo
  {author} {\bibfnamefont {S.}~\bibnamefont {Tarucha}}, \ and\ \bibinfo
  {author} {\bibfnamefont {L.~P.}\ \bibnamefont {Kouwenhoven}},\ }\href
  {\doibase 10.1038/35015509} {\bibfield  {journal} {\bibinfo  {journal}
  {Nature}\ }\textbf {\bibinfo {volume} {405}},\ \bibinfo {pages} {764}
  (\bibinfo {year} {2000})}\BibitemShut {NoStop}%
\bibitem [{\citenamefont {Petit}\ \emph {et~al.}(2014)\citenamefont {Petit},
  \citenamefont {Feuillet-Palma}, \citenamefont {Della~Rocca},\ and\
  \citenamefont {Lafarge}}]{Petit2014}%
  \BibitemOpen
  \bibfield  {author} {\bibinfo {author} {\bibfnamefont {P.}~\bibnamefont
  {Petit}}, \bibinfo {author} {\bibfnamefont {C.}~\bibnamefont
  {Feuillet-Palma}}, \bibinfo {author} {\bibfnamefont {M.~L.}\ \bibnamefont
  {Della~Rocca}}, \ and\ \bibinfo {author} {\bibfnamefont {P.}~\bibnamefont
  {Lafarge}},\ }\href {\doibase 10.1103/PhysRevB.89.115432} {\bibfield
  {journal} {\bibinfo  {journal} {Phys. Rev. B}\ }\textbf {\bibinfo {volume}
  {89}},\ \bibinfo {pages} {115432} (\bibinfo {year} {2014})}\BibitemShut
  {NoStop}%
\end{thebibliography}%
\bibliographystyle{apsrev4-1}

\end{document}